\documentclass[fleqn,usenatbib]{mnras}
\usepackage[T1]{fontenc}

\DeclareRobustCommand{\VAN}[3]{#2}
\let\VANthebibliography\thebibliography
\def\thebibliography{\DeclareRobustCommand{\VAN}[3]{##3}\VANthebibliography}

\usepackage{graphicx}
\usepackage{amsmath}
\usepackage{booktabs}
\usepackage{multirow}
\usepackage{xcolor}
\usepackage{tikz}
\usetikzlibrary{decorations.pathmorphing,arrows.meta,calc}
\usepackage{placeins}
\usepackage{balance}

\newcommand{\hmol}{H$_2$}
\newcommand{\hatom}{H}
\newcommand{\nH}{n_{\rm H}}
\newcommand{\Tdust}{T_{\rm dust}}
\newcommand{\Tgas}{T_{\rm gas}}
\newcommand{\Rfrate}{R_{{\rm H}_2}}
\newcommand{\kf}{k_f}
\newcommand{\cmcube}{cm$^{-3}$}
\newcommand{\persec}{s$^{-1}$}
\newcommand{\fchem}{f_{\rm chem}}
\newcommand{\PER}{P_{\rm ER}}
\newcommand{\sigER}{\sigma_{\rm ER}}

\usepackage{hyperref}

\title{Catalytic formation of H$_2$ on carbonaceous dust grains - implications for interstellar observations}

\author[A. Das, J. Gao, D. Vrinceanu, \& H. R. Sadeghpour]{
Aryav Das,$^{1}$\thanks{E-mail: aryav30das@gmail.com (AD); gaojulia01@gmail.com (JG)}\thanks{These authors contributed equally to this work.}
Julia Gao,$^{2}$\footnotemark[1]\footnotemark[2]
Daniel Vrinceanu$^{3}$ and
H. R. Sadeghpour$^{4}$
\\
$^{1}$Park Tudor School, 7200 N College Ave, Indianapolis, IN 46240, USA\\
$^{2}$Fairview High School, 1515 Greenbriar Blvd, Boulder, CO 80305, USA\\
$^{3}$Department of Physics, Texas Southern University, 3100 Cleburne Street, Houston, TX 77004, USA \\
$^{4}$ITAMP, Center for Astrophysics $|$ Harvard \& Smithsonian, 60 Garden Street, Cambridge, MA 02138, USA
}

\date{Accepted XXX. Received YYY; in original form ZZZ}
\pubyear{2026}

\begin{document}
\label{firstpage}
\pagerange{\pageref{firstpage}--\pageref{lastpage}}
\maketitle

\begin{abstract}
We use kinetic Monte Carlo (KMC) simulations to study molecular hydrogen formation on carbonaceous dust grain surfaces, validated against recent laboratory measurements of H$_2$ formation on coronene films at temperatures from 10 to 250\,K. The model uses a three-dimensional amorphous carbon lattice with heterogeneous physisorption ($45 \pm 5$~meV) and chemisorption ($1.75 \pm 0.25$~eV) sites, and tracks both Langmuir--Hinshelwood (LH) and Eley--Rideal (ER) formation channels within a stochastic Gillespie event-driven framework. The model reproduces the measured efficiency curve within the experimental uncertainties, including the isothermal (constant surface temperature) measurements at 100--250~K. The simulations correctly describe the phase boundary between the LH and ER driven processes as functions of grain temperature and the observed crossover. Under interstellar medium conditions, 10--250~K and $\nH = 10$--$10^4$~\cmcube, the model predicts three distinct regimes for the formation efficiency $\epsilon$, the fraction of impinging H atoms released as H$_2$. At 10\,K diffusion is slow and $\epsilon \approx 0.06$. Between 20~K and 80~K, LH dominates and $\epsilon \approx 0.28$. Above 150~K, an ER plateau at $\epsilon = 0.19$ is sustained by chemisorption-trapped \hatom\ atoms. The LH-to-ER crossover occurs between 100 and 120~K. At 100~K we observe a 16\% density-dependent stochastic enhancement, which rate-equation models cannot capture.
At $\Tdust = 60$~K, $\nH = 10^3$~\cmcube\ we find the ratio of H$_2$ formation to free-fall time $t_{{\rm H}_2}/t_{\rm ff} \approx 0.93$, so dust-catalysed \hmol\ chemistry can keep pace with gravitational collapse in high-redshift star-forming environments.
\end{abstract}


%%%%%%%%%%%%%%%%%%%%%%%%%%%%%%%%%%%%%%%%%%
\section{Introduction}
\label{sec:intro}

Molecular hydrogen is the most abundant molecule in the Universe and is central to the chemistry, thermodynamics, and evolution of interstellar environments. \hmol\ is the dominant coolant in primordial and low-metallicity gas, so it controls the thermal balance of collapsing clouds and the conditions under which stars form \citep{Gould1963}. In the contemporary ISM, \hmol\ self-shielding controls the extent of photodissociation regions and the boundary between atomic and molecular phases \citep{Hollenbach1999}. \hmol\ does not form efficiently in the gas phase under typical interstellar conditions. Gas-phase three-body recombination is negligible at densities of $10^2$--$10^4$~\cmcube, and radiative association of two \hatom\ atoms is strongly forbidden by quantum mechanical selection rules, with a rate coefficient of only $\sim 10^{-23}$~cm$^3$~\persec\ at 100~K \citep{Gould1963}. Grain-surface catalysis has long been recognised as the primary pathway for \hmol\ formation in the ISM \citep{Hollenbach1971}; \citet{Vidali2013} and \citet{Wakelam2017} review the field.

Interstellar dust grains act as both collectors and catalysts. Atomic hydrogen accretes on to their surfaces, migrates across potential wells, and recombines to form \hmol\ before being released back into the gas phase. The efficiency of this process depends on grain material, surface morphology, temperature, and irradiation environment. Two principal recombination mechanisms have been identified. The Langmuir--Hinshelwood (LH) mechanism involves two adsorbed \hatom\ atoms that diffuse thermally (or via quantum tunnelling) across the grain surface until they meet and recombine \citep{Hollenbach1971,Biham2001}. LH dominates at low temperatures where surface residence times are long and thermal mobility is sufficient. The Eley--Rideal (ER) mechanism bypasses diffusion: a gas-phase \hatom\ atom reacts directly with a surface-bound \hatom\ atom upon arrival \citep{Harris1991,Cazaux2002}. ER becomes significant when surface residence times are long, as in deep chemisorption wells, but diffusion is suppressed by high desorption rates from physisorption sites.

%\subsection{Carbonaceous grains: the underexplored substrate}
%\label{sec:intro_carbon}

The interstellar dust population divides into silicate and carbonaceous families that together account for essentially all of the refractory mass in the ISM \citep{MRN1977,Draine1984}. Most studies of \hmol\ formation have focused on silicate grains, whose laboratory analogues (olivine in particular) are well characterised with measured binding energies, diffusion barriers, and temperature-programmed desorption (TPD) kinetics \citep{Pirronello1997,Pirronello1999,Katz1999}. Many astrochemical networks adopt silicate-derived parameters as the default substrate for grain-surface chemistry.

Carbonaceous grains (graphite-like structures, amorphous carbon, hydrogenated amorphous carbon, and polycyclic aromatic hydrocarbons) have very different surface chemistries. Their surfaces sustain both physisorption sites with shallow potential wells ($E_{\rm phys} \sim 30$--70~meV; \citealt{Ghio1980}) and chemisorption traps with deep binding energies ($E_{\rm chem} \sim 0.7$--2.0~eV; \citealt{Sha2002,Rauls2008}). With two binding regimes on the same surface, \hmol\ formation can proceed over an extended temperature window. Physisorbed atoms provide the mobile reactant pool at low temperatures, while chemisorbed atoms serve as a stable reservoir for ER reactions at elevated temperatures. Carbonaceous grains also respond more strongly to ultraviolet irradiation, which can introduce or heal surface defects, modify adsorption site populations, and alter catalytic efficiency dynamically \citep{Jones2015}.

\citet{Grieco2023} used the FORMOLISM apparatus to measure \hmol\ recombination efficiency on coronene films from 10 to 250~K. They reported efficiencies of approximately 20~per~cent across the entire 100--250~K range. This was the first direct evidence that carbonaceous surfaces catalyse \hmol\ formation at temperatures well above the $\sim$20~K limit previously assumed for physisorption-dominated chemistry. The result has direct implications for the \hmol\ budget of warm interstellar environments, including photodissociation regions (PDRs; \citealt{Habart2004}) and high-redshift galaxies where the cosmic microwave background sets a temperature floor of 20--30~K at $z \sim 6$--10.

%\subsection{Previous simulation work}
%\label{sec:intro_prior}

Theoretical treatments of \hmol\ formation on grain surfaces span a hierarchy of approaches. Rate-equation models \citep{Cazaux2002,Cazaux2004} approximate surface coverage and reaction rates using averaged quantities and are widely used in astrochemical networks. The \citet{Cazaux2004} prescription, corrected by an erratum in \citet{Cazaux2010erratum} that increased efficiencies by a factor of $\sim$3.5 above 25~K through corrected transmission coefficients, remains the standard analytic baseline. Rate equations fail to capture stochastic fluctuations that dominate at low surface coverages \citep{Biham2001,Green2001}.

Master and moment equations approaches include fluctuations statistically but often require simplifying assumptions about surface geometry \citep{Lohmar2006,Lohmar2009}. More detailed studies use kinetic Monte Carlo (KMC) methods, which track individual adsorption, diffusion, and reaction events in continuous time. KMC handles surface heterogeneity, spatial correlations, and rare-event dynamics directly, and so gives a more realistic picture of molecule formation than rate equation or moment methods \citep[see][for a review of stochastic surface-chemistry methods]{Cuppen2017}.

The first microscopic KMC application to interstellar \hmol\ formation was by \citet{Chang2005}, who used a continuous-time random-walk (CTRW) algorithm to study back-diffusion and site-specific rates on olivine and amorphous carbon at 6--30~K. They showed that inhomogeneity of diffusion and desorption rates has a stronger effect on the temperature window for efficient \hmol\ formation than back-diffusion alone. \citet{Cuppen2005} extended this approach to study the effect of surface roughness on both olivine and amorphous carbon, showing that even minimal roughness raises the temperature of high efficiency ($>$50~per~cent) from 9~K to over 16~K for olivine and from 16~K to almost 30~K for amorphous carbon. \citet{Cuppen2006} added stochastic grain heating from UV photon absorption and showed that for grains smaller than 0.005~$\mu$m, temperature fluctuations exceeding 30~K occur and significantly affect LH efficiency.

\citet{Iqbal2012} applied CTRW Monte Carlo with both physisorption and chemisorption on olivine and amorphous carbon surfaces across 5--825~K, with validation against the TPD data of \citet{Pirronello1999} and \citet{Katz1999}. Their work showed efficient \hmol\ formation up to several hundred kelvin when chemisorption is included, but predates the Grieco experiment by over a decade and was validated against different experimental data.

\citet{Satonkin2025} recently presented an off-lattice Monte Carlo model on rough carbonaceous surfaces covering 5--35~K. Their model uses physisorption only (no chemisorption), with both thermal diffusion and quantum tunnelling. They found that surface roughness produces a wide, Gaussian-like distribution of binding energies (spanning a factor of 3 from minimum to maximum) and that thermal hopping dominates over tunnelling above 10~K on their model surfaces. Their model has no experimental validation.

%\subsection{Scope of this work}
%\label{sec:intro_scope}

\begin{figure}
\includegraphics[width=\columnwidth]{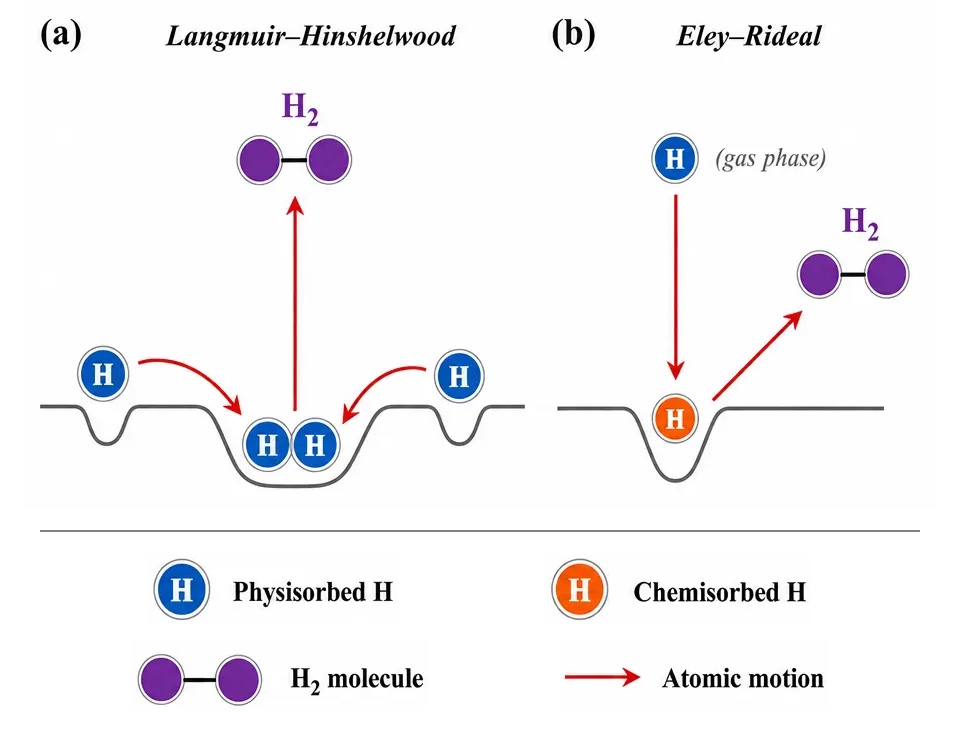}
\caption{Schematic of the two \hmol\ formation pathways treated in this work.
(a) Langmuir--Hinshelwood: two physisorbed H atoms diffuse across the surface
until they meet and recombine. (b) Eley--Rideal: an incoming gas-phase H atom
reacts directly with a chemisorbed H atom, without requiring diffusion.}
\label{fig:schematic}
\end{figure}

\begin{table*}
\centering
\caption{Comparison of published KMC/MC simulations of \hmol\ formation on grain surfaces. Only the present work validates against the high-temperature ($T > 100$~K) experimental data of \citet{Grieco2023}.}
\label{tab:comparison}
\begin{tabular}{lccccc}
\toprule
Study & Method & $T$ range (K) & Chemisorption & Expt.\ validation & Grain type \\
\midrule
\citet{Chang2005} & CTRW lattice & 6--30 & No & \citet{Biham2001} & Olivine, a-C \\
\citet{Cuppen2005} & CTRW lattice & 6--30 & No & \citet{Pirronello1999} TPD & Olivine, a-C \\
\citet{Cuppen2006} & CTRW + stoch.\ heating & 6--30 (fluct.) & No & --- & Olivine \\
\citet{Iqbal2012} & CTRW lattice & 5--825 & Yes & \citet{Pirronello1999} TPD & Olivine, a-C \\
\citet{Satonkin2025} & Off-lattice MC & 5--35 & No & --- & a-C (rough) \\
This work & Gillespie lattice & 10--250 & Yes & \citet{Grieco2023} & Coronene/a-C \\
\bottomrule
\end{tabular}
\end{table*}

We introduce a Gillespie-algorithm lattice KMC of a three-dimensional amorphous carbon grain incorporating both physisorption and chemisorption sites, compare the KMC simulation results with the Grieco laboratory data, and predict \hmol\ formation rates under the ISM conditions across 10--250~K and $\nH = 10$--$10^4$~\cmcube. A lattice representation permits explicit site-resolved binding-energy heterogeneity, periodic surface boundary conditions, and tractable enumeration of nearest-neighbour encounters at ensemble sizes large enough for converged statistics; site-to-site disorder is captured by drawing binding energies from measured distributions rather than from off-lattice geometry (cf.\ \citealt{Satonkin2025}). We quantify the LH-to-ER mechanistic crossover, identify a density-dependent stochastic enhancement at the transition temperature, compare with the \citet{Cazaux2010erratum} analytic rate prescription, integrate over the MRN grain-size distribution \citep{MRN1977}, and compute \hmol\ formation time-scales relevant to high-redshift star formation.

The salient features of this work are:

\begin{enumerate}
    \item Comparison of temperature-dependent H$_2$ formation with the laboratory measurements of  \citet{Grieco2023}(Section~\ref{sec:grieco}),
    \item Identification of three distinct ISM regimes: a diffusion-limited cold regime at 10~K, a peak LH-dominated window from 20--80~K with $\epsilon \approx 0.28$, and a stable ER-driven plateau above 150~K with $\epsilon = 0.190$ (Section~\ref{sec:epsT}),
    \item A phase crossover from LH to ER at the boundary, T=100--120~K, accompanied by a density-dependent stochastic enhancement at the transition temperatures; a feature previously unexplained from rate-equation models (Sections~\ref{sec:mechanism} and \ref{sec:density}),
    \item A volumetric \hmol\ formation rate consistent with the canonical \citet{Jura1975} value at the peak efficiency, with an MRN grain-size integration boost of roughly 30\%  in the warm regime (Section~\ref{sec:mrn}),
    \item A large formation rate of \hmol\ at $\Tdust = 60$~K, so dust-catalysed \hmol\ chemistry can keep pace with gravitational collapse under conditions relevant to high-redshift star formation (Section~\ref{sec:timescales}),
    \item A four-parameter analytic fit and metallicity-scaled rate coefficient for direct use in astrochemical networks (Section~\ref{sec:fit}, equations~\ref{eq:fit} and \ref{eq:metallicity}).
\end{enumerate}

%%%%%%%%%%%%%%%%%%%%%%%%%%%%%%%%%%%%%%%%%%
\section{Detailed KMC Simulation Protocol}
\label{sec:model}

Our methodology combines a detailed surface morphology model with explicit treatment of microscopic processes within a Gillespie event-driven kinetic Monte Carlo framework. We begin by constructing the grain lattice (Section~\ref{sec:grain}), describe the physical and chemical processes (Section~\ref{sec:processes}), the KMC algorithm (Section~\ref{sec:kmc}), and the validation protocol (Section~\ref{sec:validation}).

\subsection{Grain structure and lattice construction}
\label{sec:grain}

The baseline grain radius is $r = 0.005\,\mu$m, corresponding to a small PAH-like carbonaceous grain at the lower end of the MRN size distribution \citep{MRN1977}. The dependence of the results on grain radius is quantified in Sections~\ref{sec:mrn} and~\ref{sec:robustness}: per-grain efficiency varies by only a few per cent across a factor of 5 in radius, and the population-level effect of the size distribution is captured by the MRN integration. The lattice is constructed on a cubic grid with a 5\,\AA\ lattice constant, i.e.\ a site area $A_{\rm site} = 25$\,\AA$^2$ (approximately the area of a single coronene ring), with periodic boundary conditions applied in the surface plane to approximate extended surface behaviour.

\begin{figure}
\includegraphics[width=\columnwidth]{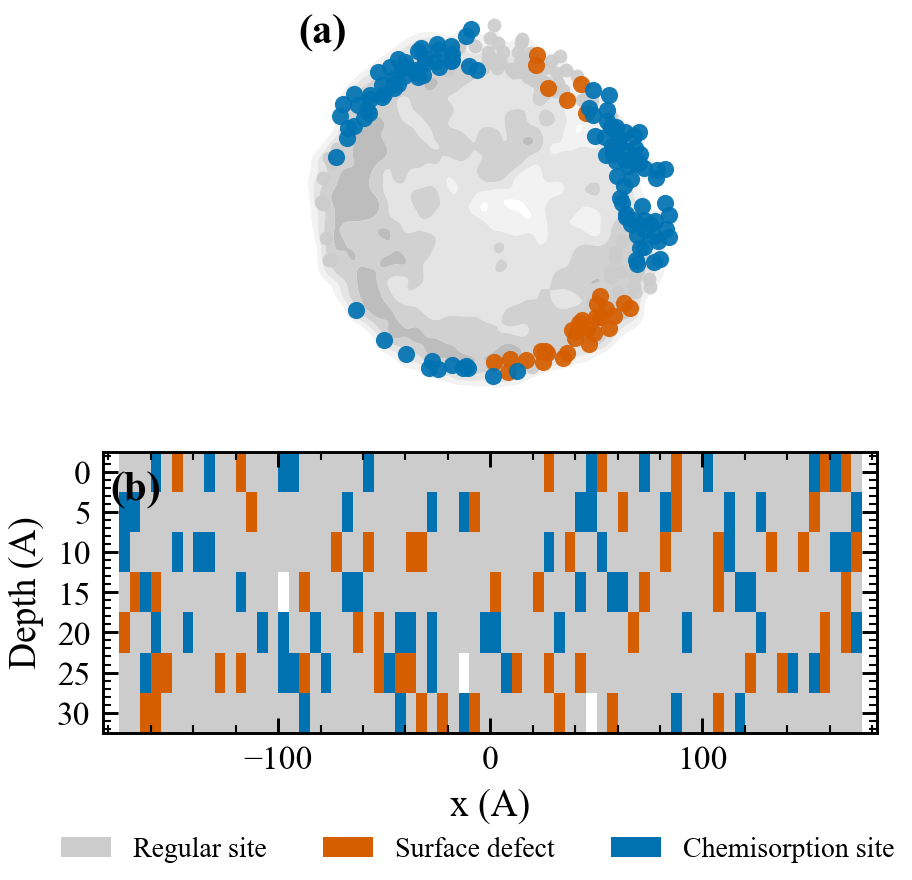}
\caption{Visualisation of the simulated amorphous carbon grain. (a) Three-dimensional projected view of the outer shell after wedge cut, with regular sites (light grey), surface defects (orange), and chemisorption-active sites (blue) coloured according to local site type. (b) Cross-sectional slice through the centre of the grain showing the depth structure of pores, defects, and chemisorption sites; white cells are pore voids, i.e.\ empty lattice sites excluded from the site network. The grain has radius 0.005~$\mu$m, porosity 0.2, surface-defect fraction 0.15, and chemisorption fraction 0.40.}
\label{fig:grain}
\end{figure}

The lattice incorporates structural heterogeneity through three features (Fig.~\ref{fig:grain}). Porosity is modelled as void inclusions within the lattice, with a void fraction of 20~per~cent for ISM configurations and 0~per~cent for laboratory validation (matching the dense coronene film in the Grieco experiment). Surface defects make up 15~per~cent of surface sites and are assigned enhanced binding geometries with intermediate binding energies. Chemisorption-active sites comprise 40~per~cent of the total surface, reflecting the high density of edge and defect sites on coronene and amorphous carbon surfaces where C--H bonds can form with binding energies of $\sim$1--2~eV. For the baseline grain this yields $\sim 1.3\times10^{3}$ surface sites, of which 60~per~cent are physisorption sites (45~per~cent regular and 15~per~cent defect-enhanced) and 40~per~cent are chemisorption-active. The occupancy of chemisorption sites is tracked dynamically throughout the simulation: at $T < 100$~K they remain largely unfilled owing to efficient LH depletion, while above 150~K the reservoir saturates (Section~\ref{sec:surface_h}). The sensitivity of $\epsilon$ to $\fchem$ is quantified in Appendix~\ref{sec:app_sensitivity}.

\begin{figure}
\includegraphics[width=\columnwidth]{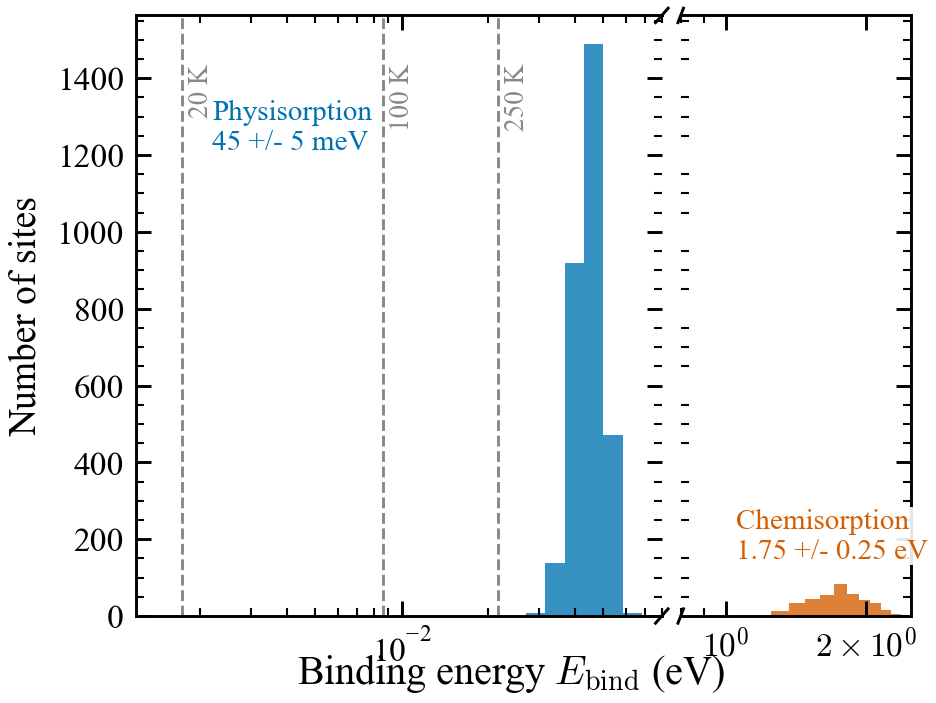}
\caption{Distribution of binding energies on the simulated grain. The two modes at 45~meV (physisorption, blue) and 1.75~eV (chemisorption, orange) are separated by nearly two orders of magnitude in energy. Vertical dashed lines mark the characteristic thermal energy $k_{\rm B}T$ at 20, 100, and 250~K, illustrating that physisorbed atoms desorb thermally above $\sim$40~K while chemisorbed atoms remain immobile across the entire temperature range considered. The broken horizontal axis emphasises the bimodal character of the binding-energy landscape.}
\label{fig:bindings}
\end{figure}

Adsorption site energetics are drawn from physically motivated Gaussian distributions. Physisorption binding energies are drawn from $\mathcal{N}(45~\text{meV},\,(5~\text{meV})^2)$, spanning approximately 30--60~meV within $3\sigma$, consistent with measured physisorption on graphite~\citep{Ghio1980}. Chemisorption binding energies are drawn from $\mathcal{N}(1.75~\text{eV},\,(0.25~\text{eV})^2)$, spanning approximately 1.0--2.5~eV within $3\sigma$, bracketing the range of \citet{Rauls2008} for armchair and zigzag edge sites on coronene.
The resulting binding-energy landscape is shown in Fig.~\ref{fig:bindings}: the physisorption and chemisorption peaks are separated by nearly two orders of magnitude in energy. This bimodal distribution is a defining feature of carbonaceous surfaces and distinguishes them from the more homogeneous binding landscapes of silicate grains. All model parameters, their values, status, and literature sources are listed in Table~\ref{tab:params}.

\begin{table*}
\centering
\caption{Model parameters, their adopted values, and literature
constraints. Parameters labelled `Adopted' are model inputs chosen
within the physically motivated ranges indicated; those labelled
`Empirical' are computational approximations without direct
literature derivation.}
\label{tab:params}
\small
\setlength{\tabcolsep}{4pt}
\begin{tabular}{@{}l c c c p{8.2cm}@{}}
\toprule
Parameter & Symbol & Value & Status & Reference / constraint \\
\midrule
Chemisorption fraction        & $\fchem$          & 0.40                              & Adopted    & Edge/defect site density on coronene \\
Physisorption binding energy  & $E_{\rm phys}$    & $45 \pm 5$~meV                    & Literature & \citet{Ghio1980}; \citet{Sha2002} \\
Chemisorption binding energy  & $E_{\rm chem}$    & $1.75 \pm 0.25$~eV                & Literature & \citet{Rauls2008} \\
Surface defect fraction       & $f_{\rm defect}$  & 0.15                              & Fixed      & Typical for a-C surfaces \\
ER cross-section              & $\sigER$          & $1.1 \times 10^{-15}$~cm$^2$      & Literature & \citet{Mennella2012}; \citet{Thrower2012} \\
ER reaction probability       & $\PER$            & 0.9                               & Adopted    & Near-barrierless ER abstraction \citep{Rauls2008, Thrower2012} \\
Sticking probability          & $S$               & 0.5                               & Fixed      & Within \citet{Hollenbach1979} range \\
Diffusion prefactor           & $\nu_{\rm diff}$  & $10^{12}$~\persec                 & Literature & \citet{Hasegawa1992}; \citet{Cazaux2004} \\
Desorption prefactor          & $\nu_{\rm des}$   & $10^{13}$~\persec                 & Literature & \citet{Hasegawa1992}; \citet{Cazaux2004} \\
Diffusion rate cap            & ---               & 200~\persec                       & Empirical  & Computational tractability \\
LH diffusion factor           & ---               & 0.5                               & Empirical  & Back-diffusion correction (Section~\ref{sec:lh_proc}) \\
LH encounter neighbours       & $z$               & 3.0                               & Empirical  & Effective coordination number \\
Grain radius                  & $r$               & 0.005~$\mu$m                      & Fixed      & Small PAH-like grain \\
Site area                     & $A_{\rm site}$    & 25~\AA$^2$                        & Fixed      & Coronene ring area \\
Porosity fraction             & ---               & 0.0 / 0.2                         & Fixed      & Validation / ISM \\
\bottomrule
\end{tabular}
\end{table*}

\subsection{Physical and chemical processes}
\label{sec:processes}

The simulation includes the adsorption, diffusion, desorption, and recombination processes most relevant to H$_2$ formation under astrophysical conditions; processes not treated explicitly (quantum-tunnelling diffusion, stochastic grain heating, UV-driven photofragmentation) are discussed in Section~\ref{sec:limitations}.

\subsubsection{Adsorption}

Two arrival-rate prescriptions are implemented, corresponding to the two applications of the model. In \emph{laboratory-validation mode}, the arrival rate is set directly to match the experimental beam flux ($\sim 0.01$ atoms per site per second, following the FORMOLISM
protocol), with a $\cos\theta$ correction for the $40^\circ$ beam incidence angle and an integrated dose of $3\times10^{15}$ atoms\,cm$^{-2}$.

In the \emph{ISM mode}, the arrival rate is derived from gas kinetics:
\begin{equation}
  F = \frac{1}{4}\, \nH\, v_{\rm th}\, A_{\rm site}\, S,
  \label{eq:flux}
\end{equation}
where $v_{\rm th} = \sqrt{8 k_{\rm B} \Tgas / (\pi m_{\rm H})}$ is the thermal velocity at a gas temperature of $\Tgas = 100$~K and $S = 0.5$ is the sticking coefficient, held constant across the temperature range; this value falls within the range of the \citet{Hollenbach1979} prescription at $\Tgas = 100$~K. Laboratory measurements indicate that the sticking coefficient declines with temperature, following
\begin{equation}
S(T) = S_0\,\frac{1 + \beta\,T/T_0}{\left(1 + T/T_0\right)^{\beta}},
\label{eq:sticking}
\end{equation}
a functional form introduced by \citet{Matar2010} for amorphous water ice and refitted for silicate surfaces by \citet{Chaabouni2012}, with $S_0 = 1$, $T_0 = 25 \pm 10$~K, and $\beta = 2.5$ for H atoms arriving as a thermal (Maxwellian) gas. The effect of replacing the constant $S$ with a temperature-dependent sticking law is quantified in Section~\ref{sec:robustness} and Appendix~\ref{sec:app_porosity}: declining sticking lowers the warm-regime rate and is the dominant systematic uncertainty on the absolute normalisation.

In either mode, each arriving atom is assigned to a physisorption or chemisorption site according to the local site type, and each arrival event simultaneously carries a probability of direct ER reaction with a surface-bound atom (Section~\ref{sec:er_proc}). The formation pathways treated by the model are illustrated schematically in Fig.~\ref{fig:schematic}.

\subsubsection{Diffusion}

Adsorbed \hatom\ atoms migrate between adjacent lattice sites at rates governed by the transition state theory:
\begin{equation}
k_{\rm diff} = \nu_{\rm diff} \exp\left(-\frac{E_{\rm diff}}{k_{\rm B} T}\right),
\label{eq:diffusion}
\end{equation}
where $\nu_{\rm diff} = 10^{12}$~\persec\ is the attempt frequency and $E_{\rm diff}$ is the site-specific diffusion barrier, set to a fixed fraction of the local binding energy, $E_{\rm diff} = 0.3\,E_{\rm bind}$, within the range commonly adopted in surface-chemistry networks \citep[e.g.][]{Hasegawa1992}. The model does not include quantum tunnelling corrections; \citet{Satonkin2025} found that thermal hopping dominates over tunnelling above 10~K on rough carbonaceous surfaces. Diffusion out of a chemisorption site is negligible at all temperatures considered here, so the mobile population consists entirely of physisorbed atoms.

\subsubsection{Langmuir--Hinshelwood recombination}
\label{sec:lh_proc}

Two adsorbed \hatom\ atoms that come to occupy neighbouring sites recombine to form \hmol, Fig.~\ref{fig:schematic}(a). For physisorbed H on carbonaceous surfaces the recombination step itself is essentially barrierless \citep{Hollenbach1971,Bron2014}, so once two mobile atoms meet, formation is assumed to proceed with unit probability. The rate-limiting step is the encounter rather than the reaction, which is why LH efficiency is governed by the competition between the hopping rate (Eq.~\ref{eq:diffusion}) and the desorption rate (Eq.~\ref{eq:desorption}).

The model implements the encounter step in two modes. In \emph{explicit-pairs mode}, used for the laboratory validation, occupied neighbouring sites are enumerated directly at each event step and recombination is evaluated pair by pair from the local site geometry. This resolves the discrete encounter statistics exactly, but requires that every diffusion hop be simulated.

In \emph{diffusion-limited mode}, used for the ISM sweep, explicit hops are computationally prohibitive: diffusion rates ($\sim 10^{12}$~\persec\ at 50~K for physisorbed atoms) vastly exceed arrival rates ($\sim 10^{-8}$~\persec\ per site), creating an extreme time-scale separation 
%in which the simulation would spend essentially all of its events on hops that change nothing. 
We therefore coarse-grain the random walks into an effective encounter rate estimated from coverage statistics,
\begin{equation}
k_{{\rm LH},i} = f_{\rm LH} \, k_{{\rm diff},i} \, z \, \theta_{\rm H},
\label{eq:lh}
\end{equation}
where $f_{\rm LH} = 0.5$ is the empirical diffusion factor, $z = 3$ is the effective coordination number, and $\theta_{\rm H} = N_{\rm H} / N_{\rm sites}$ is the fractional surface coverage. The product $k_{{\rm diff},i}\,z\,\theta_{\rm H}$ is the rate at which atom $i$ hops into a site adjacent to another occupied site; $f_{\rm LH}$ absorbs the reduction in encounter probability caused by back-diffusion, i.e.\ by atoms revisiting sites they have already sampled instead of finding new partners \citep{Chang2005,Lohmar2006}. A computational rate cap of 200~\persec\ is applied to individual diffusion events for numerical stability. The two modes agree to within 8--9~per~cent at 50--80~K and diverge by 25~per~cent at 20~K, where discrete encounter geometry matters most; this comparison is quantified in Section~\ref{sec:lh_consistency} and Appendix~\ref{sec:app_lh_mode}.

Recombination through either channel is strongly exothermic, releasing 4.48~eV per \hmol. The product molecule is only weakly bound to the surface: formed \hmol\ carries a binding energy of $E_{{\rm H}_2} = 51$~meV and a temperature-dependent sticking probability with a transition at 20~K, so that above $\sim$20~K newly formed molecules are promptly released to the gas phase and cannot block the site at which they formed. The model does not track the partitioning of the 4.48~eV between translation, rotation, vibration, and the grain lattice, which sets the internal-state distribution of the released molecule \citep[see][]{Wakelam2017} but does not affect the formation efficiency computed here.

\subsubsection{Desorption}
\label{sec:des_proc}

Thermal desorption follows Arrhenius kinetics:
\begin{equation}
k_{\rm des} = \nu_{\rm des} \exp\left(-\frac{E_{\rm bind}}{k_{\rm B} T}\right),
\label{eq:desorption}
\end{equation}
with $\nu_{\rm des} = 10^{13}$~\persec. This value follows from the vibrational frequency of the adsorbed atom in its surface well, $\nu = (2 n_s E_{\rm bind} / \pi^2 m)^{1/2}$ for a site density $n_s \approx 10^{15}$~cm$^{-2}$, which gives $3 \times 10^{12}$--$2 \times 10^{13}$~\persec\ across our binding-energy range; we adopt the standard literature value throughout \citep{Hasegawa1992,Cazaux2004}. For physisorbed atoms ($E_{\rm bind} = 45$~meV), the desorption time-scale falls exponentially from $\sim$20~ms at 20~K, to $\sim$50~ns at 40~K, and $\sim$70~ps at 80~K, causing rapid turnover of the physisorbed population. This terminates the LH channel, since it competes directly with diffusion for every physisorbed atom. Chemisorbed atoms ($E_{\rm bind} = 1.75$~eV) are effectively non-desorbing at all temperatures considered ($T \leq 250$~K), creating a stable surface reservoir.

UV photodesorption is included for $G_0 > 0$, where $G_0$ is the far-ultraviolet field strength in units of the standard interstellar radiation field, as an additional Gillespie event for each surface-bound H atom with rate
\begin{equation}
k_{\rm photo} = Y_{\rm UV} \sigma_{\rm UV}  G_0  F_{\rm Draine},
\label{eq:photo}
\end{equation}
where $Y_{\rm UV} = 0.1$ is the photodesorption yield per absorbed photon (an upper-end value; cf.\ the ice photodesorption yields of \citealt{Oberg2009}), $\sigma_{\rm UV} = 10^{-17}$~cm$^{2}$ is the UV absorption cross-section per adsorbed H atom, and $F_{\rm Draine} = 10^{8}$~photons~cm$^{-2}$~\persec\ is the standard interstellar UV photon flux. For $G_0 = 1$ this gives $k_{\rm photo} \approx 10^{-10}$~s$^{-1}$ per adsorbed atom. This event competes with thermal desorption and recombination in the Gillespie selection step. All main ISM results use $G_0 = 0$; the UV sweep is presented in Section~\ref{sec:uv}.

\subsubsection{Eley--Rideal recombination}
\label{sec:er_proc}

The ER channel bypasses the encounter step of Section~\ref{sec:lh_proc} entirely: the reaction occurs at the moment of arrival, so it does not require the incoming atom to thermalise, adsorb, or diffuse, Fig.~\ref{fig:schematic}(b). On each arrival event, the incoming \hatom\ atom has a probability of reacting directly with a surface-bound atom:
\begin{equation}
P_{\rm ER,arrival} = \frac{\sigER \times \PER \times N_{\rm surface}}{A_{\rm grain}},
\label{eq:er}
\end{equation}
where $\sigER = 1.1 \times 10^{-15}$~cm$^2$ is the adopted ER cross-section, based on experiments on coronene films and superhydrogenated PAHs \citep{Mennella2012,Thrower2012}, $\PER = 0.9$ is the reaction probability, and $N_{\rm surface} / A_{\rm grain}$ is the surface density of available reaction partners. This channel acts on both physisorbed and chemisorbed atoms. The KMC simulations at higher temperature reach a plateau in Fig.~\ref{fig:epsilon}(a), nicely corroborating the validity of Eq.~\ref{eq:er}.

\subsection{Gillespie algorithm}
\label{sec:kmc}

The time evolution is modelled using the Gillespie direct method \citep{Gillespie1977}. At each step, all possible events are assigned rates, a random event is selected with probability proportional to its rate, and the simulation clock is advanced by $\Delta t = -\ln(r)/k_{\rm tot}$, where $r \in (0,1)$ is a uniform random number and $k_{\rm tot} = \sum_j k_j$ is the sum of all event rates.

Each condition consists of 20 independent ensemble realisations, each processing 5\,000 burn-in arrivals followed by 20\,000 measured arrivals. The full ISM sweep covers 15 temperatures (10--250~K), 4 densities (10--$10^4$~\cmcube), and 4 UV levels ($G_0 = 0, 1, 10, 100$), yielding 240 conditions and 4\,800 total KMC realisations.

Formation efficiency is:
\begin{equation}
\epsilon = \frac{2 N_{{\rm H}_2}}{N_{\rm impinging}}.
\label{eq:epsilon}
\end{equation}
The efficiency is evaluated over 20\,000 measured arrivals that follow the burn-in phase, i.e.\ once the surface population has reached a statistical steady state, and is reported as the ensemble mean over the 20 independent realisations. The KMC is a stochastic method, so repeated realisations scatter about the ensemble mean; this scatter is propagated as the 95~per~cent confidence intervals quoted throughout and shown as shaded bands in the figures. The full distributions at the mechanistic transition are characterised in Appendix~\ref{sec:app_transition}.

A representative single realisation, showing the approach to steady state and the relative weights of the individual event channels, is given in Appendix~\ref{sec:app_trajectory} (Fig.~\ref{fig:trajectory}).

\subsection{Experimental validation protocol}
\label{sec:validation}

The model is validated against the \citet{Grieco2023} measurements, which span 10--250~K. Seven of the twenty points were taken under isothermal conditions, i.e.\ at fixed surface temperature during exposure (the DED protocol), spanning 20--250~K; the remaining thirteen derive from the TPDED temperature ramp. The experimental protocol is reproduced in the simulation: $40^\circ$ beam incidence, an integrated dose of $3 \times 10^{15}$~atoms~cm$^{-2}$, and zero porosity. The simulated target is the same three-dimensional lattice grain used throughout rather than a separate planar slab; with zero porosity and periodic surface boundary conditions its surface is locally equivalent to the dense, flat coronene film of the experiment, and the beam geometry enters through the $\cos\theta$ flux correction.

%%%%%%%%%%%%%%%%%%%%%%%%%%%%%%%%%%%%%%%%%%
\section{Results}
\label{sec:results}

\subsection{Laboratory measurement validation}
\label{sec:grieco}

\begin{figure*}
\includegraphics[width=\textwidth]{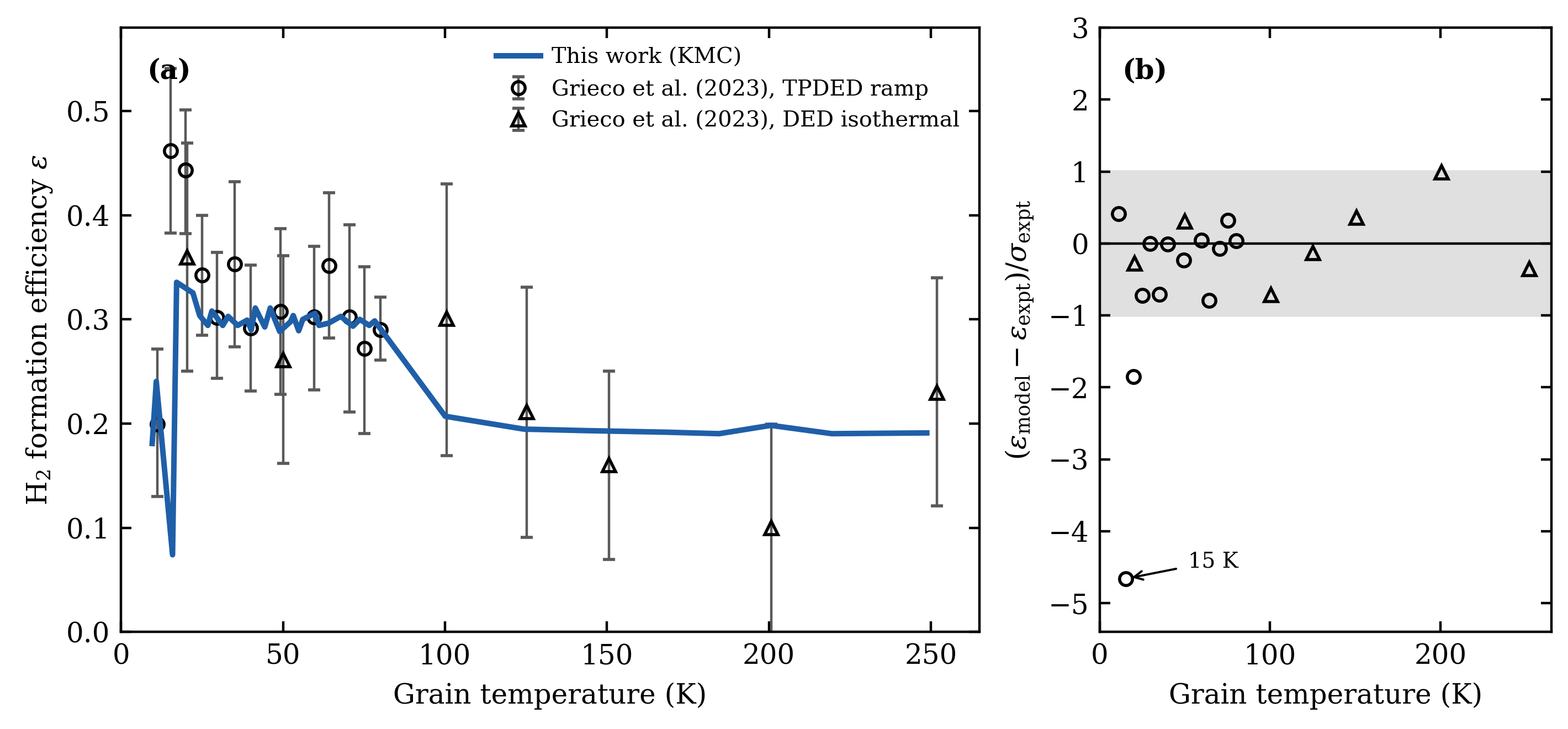}
\caption{Comparison of the KMC model with the experimental \hmol\ formation efficiencies of \citet{Grieco2023} on coronene films. (a) The solid blue curve is the KMC ensemble mean over 20 realisations per point; the 95~per~cent confidence interval ($\pm 0.001$--$0.002$) is narrower than the plotted line width. Open circles and triangles with error bars are the experimental TPDED-ramp and isothermal DED data, respectively. (b) Residuals normalised by the experimental uncertainty $\sigma_{\rm expt}$; the grey band marks $\pm 1\sigma$. Eighteen of the twenty measurements fall within $1\sigma$; the 15~K residual is discussed in Section~\ref{sec:grieco}. The simulation shows that the phase boundary near $T_p \simeq 100$~K in the measured \hmol\ formation efficiency arises from competing mechanisms: below $T_p$ formation proceeds through physisorbed H atoms (the LH channel), while above $T_p$ it crosses over to chemisorbed H atoms (the ER channel); see also Fig.~\ref{fig:schematic}.
}
\label{fig:grieco_overlay}
\end{figure*}

The KMC reproduces the experimental efficiency curve of \citet{Grieco2023}, including all seven isothermal points (Fig.~\ref{fig:grieco_overlay}). The model itself carries a stochastic uncertainty: the ensemble mean at each temperature has a 95~per~cent confidence interval of $\pm 0.001$--$0.002$ (20 realisations per point), which is narrower than the plotted line width in Fig.~\ref{fig:grieco_overlay}(a). This is an order of magnitude tighter than the experimental error bars, so the comparison is limited by the measurements. The normalised residuals in Fig.~\ref{fig:grieco_overlay}(b) show that 18 of the 20 measurements agree with the model to within $1\sigma_{\rm expt}$, including all seven isothermal points; the two exceptions are the TPDED points at 15 and 20~K, and only the 15~K point lies beyond $2\sigma_{\rm expt}$.

Beyond the statistical agreement, Fig.~\ref{fig:grieco_overlay} reveals the central physical result of this work: a phase change in the formation mechanism near $T_p \simeq 100$\,K. The efficiency curve consists of two distinct plateaux rather than a single smooth decline: an LH-dominated regime below $T_p$, fed by mobile physisorbed atoms, and an ER-dominated regime above it, sustained by the chemisorbed reservoir, joined by a sharp transition set by the collapse of the physisorbed residence time. The experimental curve of \citet{Grieco2023} traces both plateaux but cannot distinguish the mechanisms; the KMC identifies the boundary and attributes each regime to its channel (Section~\ref{sec:mechanism}).

The two regimes differ in more than which channel dominates. Below $T_p$ the surface is mobile and sparsely populated: physisorbed atoms supply the reactants, diffusion sets the rate, the inventory sits at 8--9 atoms because recombination removes atoms about as fast as they arrive, and $\epsilon$ depends on gas density through the instantaneous coverage. Above $T_p$, by contrast, the surface is static and saturated: the reactants are now chemisorbed, the arrival of a gas-phase atom sets the rate, the inventory is pinned near 212 atoms once the chemisorption sites fill, and $\epsilon$ no longer varies with density. What drives the switch is the physisorption residence time $\tau_{\rm des}(T)$, which falls exponentially with temperature and crosses the diffusion time-scale within a narrow interval; no parameter in the model changes discontinuously. Channel fraction (Section~\ref{sec:mechanism}), density sensitivity (Section~\ref{sec:density}) and surface occupancy (Section~\ref{sec:surface_h}) all change character at the same temperature, which is why we treat $T_p$ as a boundary between regimes rather than a feature of one curve.

The one measurement the model fails to reproduce is the 15~K TPDED point, where the simulated efficiency drops to $\epsilon = 0.094$ against a measured $\epsilon = 0.462 \pm 0.079$, a residual of $-4.7\sigma_{\rm expt}$ (Fig.~\ref{fig:grieco_overlay}b). The 11~K point agrees to within $0.4\sigma_{\rm expt}$, so this is a narrow suppression dip between $\sim$12 and 17~K rather than a general failure at low temperature.

The dip falls at the cold edge of the diffusion-limited window, where thermal hopping is still too slow to bring physisorbed atoms together but the residence time has already begun to shorten. Since the model uses pure Arrhenius hopping with no quantum tunnelling (Section~\ref{sec:limitations}), it underestimates mobility in exactly this interval and the LH channel is starved of encounters. Tunnelling of physisorbed \hatom\ would sustain migration below the thermal-hopping threshold \citep{Cazaux2004,Satonkin2025} and fill the dip in. We attribute the 15~K residual to this omitted channel. It lies far below the LH--ER transition and does not propagate into the warm-regime results, which are carried by ER and do not depend on surface mobility. The sensitivity of $\epsilon$ to the beam dissociation efficiency $\tau$ was tested at $\tau = 0.5$ and 0.9, yielding negligible differences ($<$2~per~cent) in the isothermal regime.

As a systematic cross-check, we also ran the validation with ISM-style transport settings (diffusion-limited mode, 20~per~cent porosity), confirming that the laboratory-configured model gives the closest agreement while the ISM-configured variant remains within acceptable bounds.

A physisorption-only comparison (chemisorption disabled) shows $\epsilon$ collapsing to near zero by 25~K, consistent with \citet{Satonkin2025} and demonstrating that chemisorption is the essential ingredient extending efficient formation to 250~K. In the physisorption-only run, LH recombination produces $\epsilon \approx 0.06$ at 10~K before declining monotonically as desorption overtakes diffusion. No ER formation occurs because chemisorbed target atoms are absent. This null experiment provides the clearest evidence that the warm-regime plateau is a direct consequence of the chemisorption reservoir rather than an artefact of model tuning.

Table~\ref{tab:grieco_summary} summarises the key validation metrics.

\begin{table}
\centering
\caption{Grieco validation summary. `Baseline' denotes the laboratory validation configuration (explicit-pairs LH mode, zero porosity, matched to the FORMOLISM protocol); `Prediction' denotes the ISM-configured variant used for the interstellar predictions, run here as a systematic cross-check.}
\label{tab:grieco_summary}
\begin{tabular}{lcc}
\toprule
Metric & Baseline & Prediction \\
\midrule
LH formation mode & pairs & diff.-limited \\
Porosity & 0.0 & 0.2 \\
$\epsilon$ plateau (150--250~K) & 0.190 & 0.190 \\
Largest residual (15~K) & $-4.7\sigma_{\rm expt}$ & --- \\
\bottomrule
\end{tabular}
\end{table}

\subsection{Temperature dependence of formation efficiency}
\label{sec:epsT}

\begin{figure*}
\includegraphics[width=\textwidth]{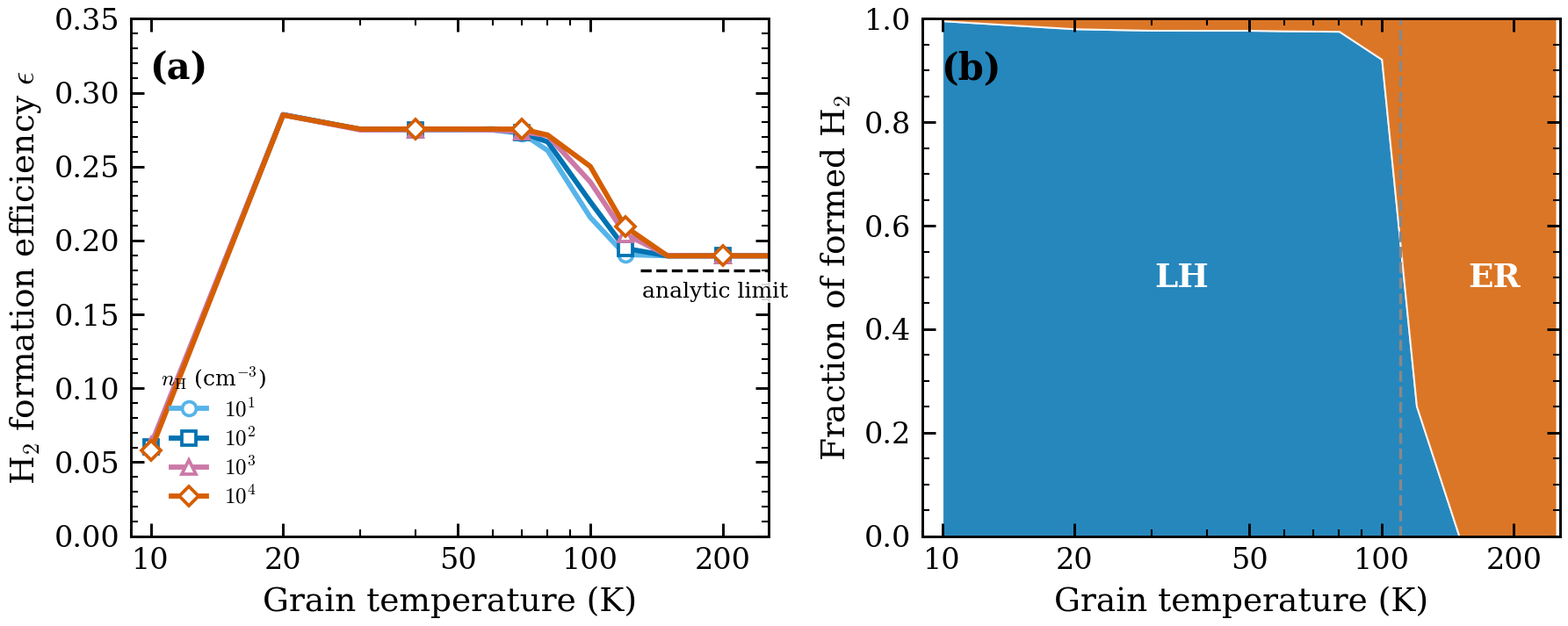}
\caption{Temperature dependence of \hmol\ formation and its mechanistic decomposition, for $G_0 = 0$; note the logarithmic temperature axes. (a) Formation efficiency at four gas densities spanning three orders of magnitude; 95~per~cent confidence intervals from 20 ensemble realisations per point are narrower than the plotted symbols. Three regimes are visible: diffusion-limited formation at 10~K, a peak Langmuir--Hinshelwood window from 20 to 80~K, and an Eley--Rideal plateau above 150~K. The density-dependent enhancement around 100~K (16~per~cent between $\nH = 10$ and $10^4$~\cmcube) is a stochastic effect absent from rate-equation treatments (Section~\ref{sec:density}). The dashed horizontal line marks the analytic high-temperature limit $\epsilon = \fchem \times \PER \times S = 0.18$: {in ER only one impinging atom is consumed per H$_2$ formed (the second reactant is pre-adsorbed), so the factor of 2 in equation~(\ref{eq:epsilon}) evaluates to unity and cancels, leaving $\epsilon_{\rm ER} = \fchem \times \PER \times S$ directly.} The efficiency stays on this limit out to 250~K instead of declining: chemisorbed atoms do not desorb over this range (Section~\ref{sec:surface_h}), so the plateau can only fall if the sticking coefficient falls, which the constant $S = 0.5$ used here prevents by construction (Fig.~\ref{fig:porosity_sticking}b). (b) Fraction of \hmol\ formed through the Langmuir--Hinshelwood (blue) and Eley--Rideal (orange) channels at $\nH = 100$~\cmcube. The ER share rises from below 10~per~cent at 100~K to 100~per~cent by 150~K as the physisorption residence time collapses; the dashed vertical line marks the crossover near 110~K.}
\label{fig:epsilon}
\end{figure*}

The ISM sweep reveals three distinct temperature regimes (Fig.~\ref{fig:epsilon}, Table~\ref{tab:rates}).

\begin{table*}
\centering
\caption{\hmol\ formation efficiency $\epsilon$, release rate $\Rfrate$, and mechanism fractions across the full ISM parameter grid ($G_0 = 0$). Each entry represents 20 independent KMC realisations of 20\,000 measured arrivals. The 95~per~cent confidence intervals on $\epsilon$ are $\pm 0.001$--0.002 throughout. Rates scale linearly with $\nH$ at all temperatures except 100--120~K where the density-dependent stochastic enhancement (Section~\ref{sec:density}) is visible.}
\label{tab:rates}
\begin{tabular}{rrcccccc}
\toprule
$T$ & $\nH$ & $\epsilon$ & $\Rfrate$ & LH & ER & $\langle N_{\rm H} \rangle$ \\
(K) & (\cmcube) & & (cm$^{-2}$~\persec) & (per~cent) & (per~cent) & (atoms) \\
\midrule
10  & 10    & 0.059 & $6.6 \times 10^3$ & 99.7 & 0.3 & --- \\
10  & 100   & 0.060 & $6.7 \times 10^4$ & 99.6 & 0.4 & --- \\
10  & 1\,000  & 0.062 & $6.9 \times 10^5$ & 99.7 & 0.3 & --- \\
10  & 10\,000 & 0.058 & $6.5 \times 10^6$ & 99.6 & 0.4 & --- \\
\midrule
20  & 10    & 0.285 & $3.2 \times 10^4$ & 98.1 & 1.9 & --- \\
20  & 100   & 0.285 & $3.2 \times 10^5$ & 98.1 & 1.9 & --- \\
20  & 1\,000  & 0.285 & $3.2 \times 10^6$ & 98.1 & 1.9 & --- \\
20  & 10\,000 & 0.285 & $3.2 \times 10^7$ & 98.1 & 1.9 & --- \\
\midrule
100 & 10    & 0.216 & $2.4 \times 10^4$ & 86.2 & 13.8 & 39 \\
100 & 100   & 0.226 & $2.5 \times 10^5$ & 92.2 & 7.8  & 25 \\
100 & 1\,000  & 0.240 & $2.7 \times 10^6$ & 95.0 & 5.0  & 13 \\
100 & 10\,000 & 0.250 & $2.8 \times 10^7$ & 96.4 & 3.6  & 12 \\
\midrule
150 & 100   & 0.190 & $2.1 \times 10^5$ & 0.0  & 100.0 & --- \\
200 & 100   & 0.190 & $2.1 \times 10^5$ & 0.0  & 100.0 & --- \\
250 & 100   & 0.190 & $2.1 \times 10^5$ & 0.0  & 100.0 & --- \\
\bottomrule
\end{tabular}
\end{table*}

At 10~K, $\epsilon \approx 0.06$: diffusion is slow, atoms land without migrating, and encounter probability is low. At 20~K, $\epsilon$ rises sharply to 0.285 as thermal activation enables LH encounters before desorption. This peak persists with slight decline through 30--80~K ($\epsilon \approx 0.275$). Above 80~K, increasing desorption of physisorbed atoms competes with diffusion, $\epsilon$ drops to 0.226 at 100~K, and the ER channel takes over. By 150~K, a plateau of $\epsilon = 0.190$ is reached and maintained through 250~K, sustained entirely by ER acting on the chemisorbed reservoir.

One might expect $\epsilon$ to fall as the grain warms from 150 to 250~K, but the simulation shows no decline at all. The ER channel draws on chemisorbed atoms whose thermal desorption time-scales exceed the age of the Universe across this entire range (Section~\ref{sec:surface_h}), so the reservoir cannot thermally deplete and the efficiency stays locked at the analytic value $\fchem \times \PER \times S$ (equation~\ref{eq:highT}). The \citet{Grieco2023} measurements likewise show no decline up to 250~K. What would eventually bend the plateau downward at higher temperatures is the sticking coefficient: the constant $S = 0.5$ adopted here suppresses that decline by construction, whereas the temperature-dependent sticking law of equation~(\ref{eq:sticking}) lowers the warm-regime efficiency (Appendix~\ref{sec:app_porosity}; Fig.~\ref{fig:porosity_sticking}b). The plateau itself is therefore a firm prediction of the model; its height depends on the sticking law.

\subsection{Mechanistic crossover from LH to ER}
\label{sec:mechanism}

Below 100~K, the LH pathway accounts for over 97~per~cent of all \hmol\ formed (Fig.~\ref{fig:epsilon}b). At 100~K, the ER fraction rises to 7.8~per~cent ($\nH = 100$~\cmcube). The crossover is remarkably sharp: between 100 and 120~K, the ER fraction jumps from 7.8 to 74.8~per~cent. By 150~K, LH collapses entirely as physisorbed atoms desorb too rapidly for diffusive encounters. Above 150~K, ER accounts for 100~per~cent of \hmol\ production through direct reaction with the chemisorbed reservoir.

The sharpness of this crossover reflects the exponential sensitivity of physisorption residence time to temperature. At 80~K, a physisorbed atom with $E_{\rm bind} = 45$~meV has a desorption time-scale of $\tau_{\rm des} = \nu_{\rm des}^{-1} \exp(E_{\rm bind}/k_{\rm B}T) \approx 10^{-10}$~s, long enough for multiple diffusion hops ($\tau_{\rm diff} \sim 10^{-12}$~s) before desorption. By 120~K, $\tau_{\rm des}$ drops to $\sim 10^{-11}$~s, allowing at most a single hop before the atom escapes. This exponential collapse of the LH window drives the rapid mechanistic switch to ER.

The crossover temperature is approximately 20~K higher than predicted by the \citet{Cazaux2004} rate-equation model. This is likely because the KMC captures back-diffusion effects (atoms returning to previously visited sites rather than finding new reaction partners) that reduce the effective LH encounter probability and postpone the point at which ER overtakes LH.

The ER fraction also shows a weak but systematic density dependence at the crossover. At 120~K, ER accounts for 74.8~per~cent of formation at $\nH = 100$~\cmcube\ but only 68~per~cent at $\nH = 10^4$~\cmcube, because higher flux sustains a marginally higher physisorbed population that extends the LH contribution. This density-dependent crossover temperature is another feature absent from rate-equation treatments.

Below $\sim$25~K the physisorption-only variant is indistinguishable from the full model: LH recombination of physisorbed H atoms alone accounts for the formation efficiency in this regime, and chemisorption plays no significant role (Fig.~\ref{fig:epsilon}b).

The mechanistic crossover has a direct experimental interpretation. The transition from LH to ER at 100--120\,K constitutes a phase change in the formation mechanism: below the transition, H$_2$ production relies on the thermally fragile physisorbed population, whereas above it, formation is carried entirely by gas-phase atoms reacting directly with the chemisorbed reservoir. This provides a microphysical explanation for the central puzzle posed by the \citet{Grieco2023} measurements: efficient H$_2$ formation ($\epsilon \approx 0.2$) persisting to 250\,K, far beyond the $\sim$20\,K ceiling expected for physisorption-dominated chemistry. \citet{Grieco2023} established this warm-regime efficiency empirically but could not isolate the responsible mechanism; our simulations show that it is the ER channel acting on chemisorption-trapped H atoms --- a reservoir that is thermally stable across the entire measured range --- that sustains the observed plateau. The persistence of H$_2$ formation on warm carbonaceous grains is therefore the expected signature of a chemisorption-fed Eley--Rideal regime rather than an anomaly.

\subsection{Density-dependent stochastic enhancement}
\label{sec:density}

At 100~K, $\epsilon$ increases from 0.216 at $\nH = 10$~\cmcube\ to 0.250 at $\nH = 10^4$~\cmcube. This is a 16~per~cent enhancement (Table~\ref{tab:density}). The density dependence vanishes below 80~K ($<$1~per~cent variation) and above 150~K (plateau set by $\fchem \times \PER \times S$).

Higher gas flux raises the instantaneous surface population, increasing LH encounter probability before physisorbed atoms desorb. This effect is intrinsically absent from rate-equation models that assume $\epsilon$ is independent of density at fixed temperature. The steady-state surface \hatom\ count decreases from 39 atoms at $\nH = 10$ to 12 at $\nH = 10^4$, reflecting more efficient depletion by the enhanced LH channel.

\begin{table}
\centering
\caption{Density-dependent formation efficiency at the LH-to-ER transition ($T = 100$~K, $G_0 = 0$). The 16~per~cent increase in $\epsilon$ from $\nH = 10$ to $10^4$~\cmcube\ is a stochastic effect absent from rate-equation models.}
\label{tab:density}
\begin{tabular}{rcccr}
\toprule
$\nH$ & $\epsilon$ & CI$_{95}$ & ER & $\langle N_{\rm H} \rangle$ \\
(\cmcube) & & & (per~cent) & (atoms) \\
\midrule
10     & 0.216 & 0.002 & 13.8 & 39 \\
100    & 0.226 & 0.001 & 7.8  & 25 \\
1\,000   & 0.240 & 0.001 & 5.0  & 13 \\
10\,000  & 0.250 & 0.001 & 3.6  & 12 \\
\bottomrule
\end{tabular}
\end{table}

\subsection{ISM formation rates and observational benchmarks}
\label{sec:rates}

\begin{figure}
\includegraphics[width=\columnwidth]{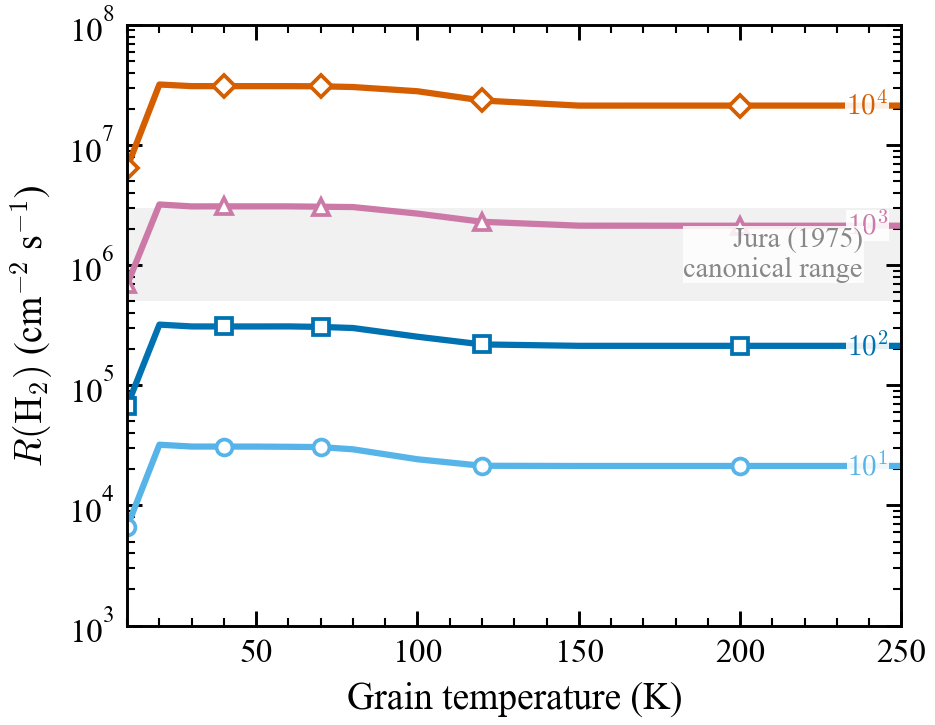}
\caption{Volumetric \hmol\ release rate $R({\rm H}_2)$ as a function of temperature for the four gas densities. The rate scales nearly linearly with $\nH$ everywhere except in the 100--120~K transition band, where the stochastic density enhancement produces a weak super-linear correction. The shaded horizontal band indicates the canonical ISM rate coefficient range $(1$--$6) \times 10^{-17}$~cm$^3$~\persec\ derived from UV absorption \citep{Jura1975,Gry2002,Wolfire2008}, converted to a per-grain rate assuming $\sigma_{\rm H} = 10^{-21}$~cm$^2$.}
\label{fig:release_rate}
\end{figure}

The release rate $\Rfrate$ scales nearly linearly with $\nH$ at fixed temperature (Fig.~\ref{fig:release_rate}, Table~\ref{tab:rates}), as expected when $\epsilon$ is approximately constant. This linear scaling holds exactly above 150~K where $\epsilon$ is density-independent, and nearly so below 80~K where the density dependence is $<$1~per~cent. At the transition temperature (100~K), the density-dependent enhancement of $\epsilon$ introduces a weak super-linear correction.

At the high-$T$ plateau ($T \geq 150$~K):
\begin{equation}
\Rfrate \approx 2.13 \times 10^4 \times \frac{\nH}{10}~~{\rm cm}^{-2}~{\rm s}^{-1}.
\end{equation}

Converting to the volumetric rate coefficient using a standard grain cross-section per \hatom\ nucleus of $\sigma_{\rm H} = 10^{-21}$~cm$^2$ \citep{Draine2003}:
\begin{equation}
\kf \approx 1.0 \times 10^{-17}~~{\rm cm}^3~{\rm s}^{-1} \quad (\Tdust = 20~{\rm K}),
\end{equation}
within a factor of 3 of the canonical observational value $\kf \approx 3 \times 10^{-17}$~cm$^3$~\persec\ derived from \textit{Copernicus} and \textit{FUSE} UV absorption measurements \citep{Jura1975,Gry2002}. The accepted range of observational determinations spans $(1$--$6) \times 10^{-17}$~cm$^3$~\persec\ \citep{Wolfire2008}. The remaining factor-of-three gap is expected for a single-grain, single-composition model: silicate grains (which contribute roughly half the total dust surface area in the ISM) are not included, nor is the full diversity of carbonaceous surface types. The MRN integration in Section~\ref{sec:mrn} partially closes this gap but cannot fully account for the missing silicate contribution.

At the peak efficiency ($T = 20$~K, $\epsilon = 0.285$), the volumetric rate reaches $\kf \approx 1.5 \times 10^{-17}$~cm$^3$~\persec\ for our single carbonaceous grain model. Including MRN integration ($1.29\times$ boost) brings this to $\kf \approx 1.9 \times 10^{-17}$~cm$^3$~\persec, within a factor of 1.6 of the canonical value from carbonaceous grains alone.

\subsection{Fitted rate coefficient for astrochemical networks}
\label{sec:fit}

To facilitate use in astrochemical models, we provide an analytic fit to the MRN-integrated efficiency as a function of grain temperature. The KMC results are well represented by a four-parameter combination of a low-$T$ Gaussian peak plus a high-$T$ analytic plateau:
\begin{equation}
\epsilon(\Tdust) = \epsilon_{\rm plateau} + (\epsilon_{\rm peak} - \epsilon_{\rm plateau})\,
\exp\left[ - \frac{(\Tdust - T_{\rm peak})^2}{2\,\sigma_T^2} \right]
\label{eq:fit}
\end{equation}
with $\epsilon_{\rm plateau} = 0.187$, $\epsilon_{\rm peak} = 0.284$, $T_{\rm peak} = 35$~K, and $\sigma_T = 50$~K (RMS residual $<$3~per~cent across 20--250~K at $\nH = 10^3$~\cmcube). Below 20~K an additional suppression factor $\exp[-(20/T)^2]$ captures the diffusion-limited collapse. Equation~(\ref{eq:fit}) is intended for fast lookup in network models such as KROME \citep{Grassi2014} or PRODIMO \citep{Woitke2009}, complementing the higher-fidelity tabulated grids in Appendix~\ref{sec:app_fullgrid}. Networks requiring per-density behaviour should interpolate the full grid in Tables~\ref{tab:rates} and \ref{tab:fullgrid}; equation~(\ref{eq:fit}) captures the $\nH$-averaged temperature dependence.

For galaxy-scale models, the volumetric rate coefficient combining equation~(\ref{eq:fit}) with the MRN integration boost is
\begin{equation}
R(T_{\rm dust}) = 3.0 \times 10^{-17}\,Z'\,\left[ \epsilon(T_{\rm dust})/\epsilon_{\rm plateau} \right]\;{\rm cm}^{3}\,{\rm s}^{-1},
\label{eq:metallicity}
\end{equation}
where $Z'$ is the metallicity in solar units \citep[following the convention of][]{Wakelam2012,Sternberg2014}. The pre-factor $3.0 \times 10^{-17}$~cm$^3$~\persec\ matches our MRN-integrated plateau rate at solar metallicity; the temperature dependence is encoded in the dimensionless ratio $\epsilon(T_{\rm dust})/\epsilon_{\rm plateau}$. This form preserves backward compatibility with the canonical \citet{Jura1975} normalisation while incorporating the carbonaceous-grain temperature dependence derived in this work.

\subsection{MRN grain-size integration}
\label{sec:mrn}

We compute the cross-section-weighted MRN-integrated rate from 12 logarithmically spaced grain-size bins spanning $a = 0.005$--0.25~$\mu$m \citep{MRN1977}, each with 20 ensemble realisations at $\nH = 10^3$~\cmcube. Each bin is weighted by $n(a) \propto a^{-3.5}$ multiplied by the grain cross-section $\pi a^2$, giving an effective weight $\propto a^{-1.5} \Delta a$.

The MRN rate boost is $1.27$--$1.30\times$ in the warm regime and $1.78\times$ at 10~K (Table~\ref{tab:mrn}). This modest correction arises because our baseline grain (0.005~$\mu$m) already sits at the small-grain peak of the MRN surface-area budget. By contrast, \citet{LeBourlot2012} found a $\sim$3.4$\times$ boost from a single 0.1~$\mu$m grain, since their baseline was near the large-grain end. The size-dependent efficiency at 10~K is notable: the smallest bins ($a = 0.005$~$\mu$m) give $\epsilon \approx 0.06$ while the largest ($a = 0.25$~$\mu$m) give $\epsilon \approx 0.27$, reflecting the longer residence times and higher surface coverage achievable on larger grains at low temperature. The size-resolved efficiencies and the cross-section weighting are shown in Fig.~\ref{fig:mrn} (Appendix~\ref{sec:app_grain}).

\begin{table}
\centering
\caption{MRN-integrated \hmol\ formation efficiency and rate boost relative to the single-grain baseline ($a = 0.005$~$\mu$m), at $\nH = 10^3$~\cmcube\ and $G_0 = 0$. The cross-section-weighted integration spans 12 grain sizes from 0.005 to 0.25~$\mu$m.}
\label{tab:mrn}
\begin{tabular}{rccc}
\toprule
$T$ & $\epsilon_{\rm MRN}$ & $\Rfrate^{\rm MRN}$ & Rate boost \\
(K) & & (cm$^{-2}$~\persec) & \\
\midrule
10  & 0.085 & $1.2 \times 10^6$ & 1.78 \\
20  & 0.284 & $4.1 \times 10^6$ & 1.29 \\
60  & 0.275 & $4.0 \times 10^6$ & 1.29 \\
100 & --- & --- & --- \\
150 & 0.187 & $2.7 \times 10^6$ & 1.27 \\
200 & 0.187 & $2.7 \times 10^6$ & 1.27 \\
250 & 0.187 & $2.7 \times 10^6$ & 1.27 \\
\bottomrule
\end{tabular}
\end{table}

\subsection{UV photodesorption suppression}
\label{sec:uv}

At $G_0 = 1$, UV photodesorption has negligible effect ($<$1~per~cent). At $G_0 = 100$, $\epsilon$ is suppressed by 28~per~cent at 200~K (from 0.190 to 0.136) by removing \hatom\ atoms from the chemisorbed reservoir. The suppression is temperature-dependent (Table~\ref{tab:uv}): at 100~K where rapid gas-phase replenishment buffers the loss, $\epsilon$ drops by only 2.7~per~cent even at $G_0 = 100$. UV-driven \hmol\ formation via PAH photofragmentation \citep{Boschman2015,Castellanos2018} is not included (competitive only at $G_0 > 10^3$).

\begin{table}
\centering
\caption{UV photodesorption suppression of formation efficiency at $\nH = 100$~\cmcube. Suppression is defined as $1 - \epsilon(G_0)/\epsilon(G_0 = 0)$.}
\label{tab:uv}
\begin{tabular}{rcccc}
\toprule
$G_0$ & $\epsilon$ (100~K) & Supp. & $\epsilon$ (200~K) & Supp. \\
\midrule
0   & 0.226 & ---  & 0.190 & --- \\
1   & 0.226 & $<$1~per~cent & 0.191 & $<$1~per~cent \\
10  & 0.225 & $<$1~per~cent & 0.188 & 1~per~cent \\
100 & 0.220 & 2.7~per~cent & 0.136 & 28~per~cent \\
\bottomrule
\end{tabular}
\end{table}

\subsection{Surface \hatom\ atom inventory}
\label{sec:surface_h}

\begin{figure}
\includegraphics[width=\columnwidth]{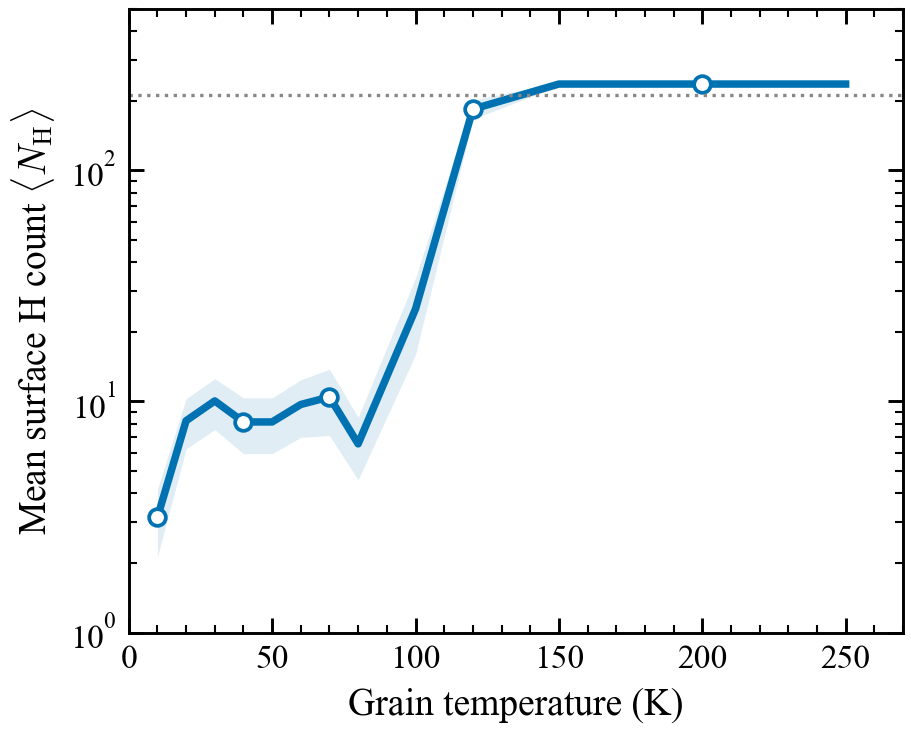}
\caption{Steady-state mean surface atomic hydrogen count as a function of grain temperature at $\nH = 100$~\cmcube\ and $G_0 = 0$, on logarithmic vertical axis. Occupancy remains at 8--9 atoms across the LH-dominated regime where rapid recombination depletes the surface, rises sharply at 100--150~K as LH weakens, and saturates at $\sim$212 atoms above 150~K (dotted horizontal line) where the chemisorption reservoir is filled. Chemisorbed atoms are effectively permanent residents at all temperatures considered.}
\label{fig:surface_h}
\end{figure}

The mechanistic transition is reflected in the steady-state surface \hatom\ population (Fig.~\ref{fig:surface_h}). At $\nH = 100$~\cmcube\ and $G_0 = 0$, the mean surface \hatom\ count remains low throughout the LH-dominated regime, holding between 8 and 9 atoms across 20--80~K as efficient recombination continuously depletes the surface inventory. The low occupancy confirms that the grain operates in the accretion limit at these temperatures: on average, fewer than two physisorbed atoms coexist on the surface at any given time, and recombination occurs rapidly upon each LH encounter.

Above 100~K, occupancy rises sharply as the declining LH channel removes fewer atoms per unit time. By 150~K, the surface population plateaus at approximately 212 atoms, all of which reside in chemisorption sites ($E_{\rm bind} = 1.75$~eV). These atoms are effectively permanent residents: at 250~K, the thermal desorption time-scale for a chemisorbed atom is $\tau_{\rm des} = \nu_{\rm des}^{-1} \exp(E_{\rm bind}/k_{\rm B}T) \approx 2 \times 10^{22}$~s, some four orders of magnitude longer than the age of the Universe. The chemisorbed population therefore accumulates monotonically until the 40~per~cent of surface sites designated as chemisorption-active are saturated, at which point the surface \hatom\ count stabilises.

This transition from a depleted surface (efficient LH) to a populated surface (stable ER reservoir) provides an independent diagnostic of the mechanistic crossover that does not depend on the formation rate itself. It also explains why the ER plateau efficiency is insensitive to gas density: the chemisorbed reservoir is saturated at all densities, so the ER rate per arrival is constant.

\subsection{High-temperature analytic limit}
\label{sec:analytic}

Above 150~K, all physisorbed atoms desorb instantly, LH ceases, and $\epsilon$ reduces to:
\begin{equation}
\epsilon_{\rm high\text{-}T} = \fchem \times \PER \times S = 0.40 \times 0.9 \times 0.5 = 0.18,
\label{eq:highT}
\end{equation}
consistent with the simulated 0.190 (the 5~per~cent offset reflects geometric corrections on the finite lattice). The identity of $\epsilon$ across 150--250~K and all densities confirms convergence to this analytic limit.

\subsection{Comparison with the Cazaux \& Tielens prescription}
\label{sec:ct10}

\begin{figure}
\includegraphics[width=\columnwidth]{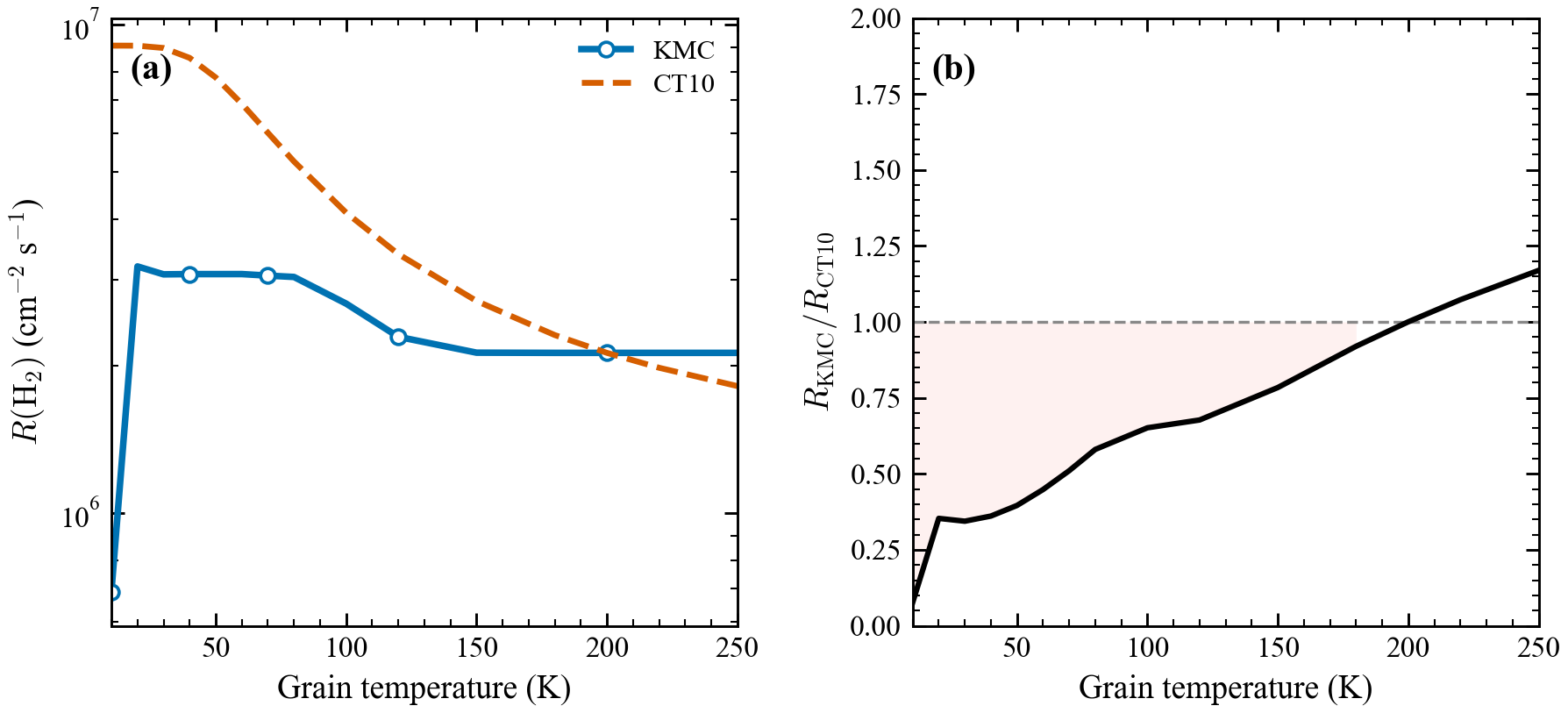}
\caption{Comparison with the erratum-corrected rate-equation prescription of \citet{Cazaux2010erratum}. (a) Volumetric \hmol\ release rate at $\nH = 10^3$~\cmcube\ from the KMC (solid blue) and CT10 (dashed orange) models. (b) Ratio $R_{\rm KMC} / R_{\rm CT10}$ as a function of temperature; the dashed grey line marks parity. The KMC predicts rates a factor of $\sim$2 lower than CT10 in the 80--150~K transition band, attributable to back-diffusion which is resolved in KMC but absent from rate equations \citep{Chang2005}. At $T > 150$~K both models converge to an ER-dominated regime.}
\label{fig:ct10}
\end{figure}

The KMC rates lie below the erratum-corrected \citet{Cazaux2010erratum} prescription by a factor of a few in the 80--150~K band (Fig.~\ref{fig:ct10}). This discrepancy has two origins. First, the CT10 model does not account for back-diffusion. Back-diffusion is the process by which a mobile \hatom\ atom returns to its original site rather than encountering a reaction partner, and the KMC resolves it explicitly. It reduces LH efficiency by factors of 2--3 relative to mean-field treatments \citep{Chang2005,Cuppen2005}. Second, the CT10 model assumes different surface parameters. Its chemisorption well depth and tunnelling barriers are derived from a two-site (physisorption + chemisorption) model with explicit quantum tunnelling between wells, whereas our model uses a heterogeneous lattice with site-specific energetics drawn from distributions.

At $T < 20$~K, the KMC and CT10 rates are in closer agreement because physisorption dominates in both models. Above 150~K, both converge to an ER-driven regime, though the absolute rate differs due to different $\fchem$ and $\PER$ assumptions. The KMC provides a stochastic correction to the CT10 rates that is most significant in the 80--150~K transition band where both mechanisms contribute and their relative weights depend on the specific surface configuration.

\subsection{Astrophysical time-scales}
\label{sec:timescales}

The validated model gives the following \hmol\ formation-to-free-fall time-scale ratios:
\begin{equation}
\frac{t_{{\rm H}_2}}{t_{\rm ff}} = \begin{cases}
0.295 & (T = 40~{\rm K},\; \nH = 10^4~{\rm cm}^{-3}) \\
0.933 & (T = 60~{\rm K},\; \nH = 10^3~{\rm cm}^{-3}) \\
3.59\phantom{0} & (T = 100~{\rm K},\; \nH = 10^2~{\rm cm}^{-3}).
\end{cases}
\end{equation}
The $t_{{\rm H}_2}$ values are computed as $t_{{\rm H}_2} = \nH / (2 \Rfrate)$, where the factor of 2 accounts for the consumption of two \hatom\ atoms per \hmol\ molecule, and $t_{\rm ff} = \sqrt{3\pi / (32 G \rho)}$ is the gravitational free-fall time at the given density assuming mean molecular weight $\mu = 1.4$.

The $T = 60$~K case sits near the critical threshold identified by \citet{Krumholz2012}, where $t_{{\rm H}_2}/t_{\rm ff} \sim 1$ and the molecular fraction at equilibrium begins to drop below 50~per~cent. Below this threshold ($T = 40$~K, $\nH = 10^4$), the gas fully converts to \hmol\ before collapse completes. Above it ($T = 100$~K, $\nH = 10^2$), the formation time-scale is 3.6 free-fall times, insufficient for molecular conversion and implying star formation proceeds in primarily atomic gas.

The conditions $T_{\rm dust} = 60$~K and $\nH = 10^3$~\cmcube\ are astrophysically plausible for dense star-forming clumps at $z \sim 6$--10, where CMB temperatures reach $T_{\rm CMB} = 2.73(1+z) \approx 19$--30~K, metallicities are $Z \sim 0.01$--$0.1~Z_\odot$, and densities in star-forming clumps reach $10^3$--$10^4$~\cmcube.

%%%%%%%%%%%%%%%%%%%%%%%%%%%%%%%%%%%%%%%%%%
\section{Model robustness}
\label{sec:robustness}

We conducted a suite of sensitivity and consistency tests to assess the robustness of the ISM predictions. Table~\ref{tab:robustness} provides a summary; full details for each test follow.

\begin{table}
\centering
\caption{Summary of model robustness checks. All tests at $G_0 = 0$ unless noted. $\Delta\epsilon$ is the change relative to the baseline configuration.}
\label{tab:robustness}
\begin{tabular}{lccr}
\toprule
Test & $T$ (K) & Variant & $\Delta\epsilon$ \\
\midrule
Porosity & 100 & 0.0 $\to$ 0.2 & $-0.5$\,\% \\
Grain size & 20 & 0.002~$\mu$m & few \% \\
Grain size & 20 & 0.010~$\mu$m & few \% \\
Sticking model & 100 & $S(T)$ & lower \\
LH mode & 20 & diff.-limited & $-25$\,\% \\
LH mode & 50 & diff.-limited & $-8$\,\% \\
LH mode & 80 & diff.-limited & $-9$\,\% \\
$\tau$ sensitivity & 200 & 0.5 vs 0.9 & $<2$\,\% \\
Transport knobs & 20--100 & varied & $\sim$ unity \\
\bottomrule
\end{tabular}
\end{table}

\subsection{Parameter sensitivity envelope}
\label{sec:sensitivity}

The two most influential model parameters ($\fchem$ and $\PER$) were swept across a $3 \times 3$ grid at four temperatures ($T = 100, 150, 200, 250$~K). The resulting $k_{\rm eff}$ envelope spans $8.09 \times 10^{-18}$ to $1.99 \times 10^{-17}$~cm$^3$~\persec\ at 100~K and $2.49 \times 10^{-18}$ to $1.75 \times 10^{-17}$ at 150--250~K. The warm-regime plateau scales nearly linearly with $\fchem \times \PER$, confirming the analytic expectation from equation~(\ref{eq:highT}). The full sensitivity grid is provided in Appendix~\ref{sec:app_sensitivity}.

\subsection{LH mode consistency}
\label{sec:lh_consistency}

The diffusion-limited approximation (equation~\ref{eq:lh}) used for ISM predictions was cross-checked against the explicit-pairs mode used for Grieco validation. At 50 and 80~K, the two modes agree to within 8--9~per~cent. At 20~K, where the LH channel is most sensitive to local encounter geometry, the diffusion-limited mode underestimates $\epsilon$ by approximately 25~per~cent. This offset is confined to the LH-dominated regime; at $T > 100$~K where ER dominates, the two modes produce identical results because the ER calculation does not depend on the LH mode.

\subsection{Porosity, grain size, and sticking}

Changing porosity from 0.0 (validation) to 0.2 (ISM) alters $\epsilon$ by only 0.5~per~cent at 100~K, confirming that the geometry change between validation and ISM configurations does not introduce significant systematic bias. Testing three grain radii (0.002, 0.005, 0.01~$\mu$m) at 20 and 150~K yields only few-per-cent shifts, consistent with the modest MRN correction factors found in Section~\ref{sec:mrn}. Replacing the constant $S = 0.5$ with an empirical temperature-dependent sticking law lowers the warm-regime rate; this provides a lower bound on the formation rate under the most pessimistic sticking assumptions.

\subsection{Transport parameter sensitivity}

Varying the LH diffusion factor from 0.3 to 0.8 and the diffusion rate cap from 50 to 500~\persec\ modulates the low-$T$ efficiency but does not overturn the warm-regime conclusions. The ER-dominated plateau is insensitive to transport parameters because ER formation occurs on arrival, bypassing diffusion entirely.

\subsection{Statistical convergence}

At 100~K with $\nH = 10^4$~\cmcube, a deep campaign of 1\,000 independent realisations confirms rapid convergence (Appendix~\ref{sec:convergence}). The SEM drops from $8.3 \times 10^{-4}$ at $N = 10$ to $9.8 \times 10^{-5}$ at $N = 1\,000$, and the ensemble mean shifts by less than 0.15~per~cent between $N = 100$ and $N = 1\,000$.

%%%%%%%%%%%%%%%%%%%%%%%%%%%%%%%%%%%%%%%%%%
\section{Discussion}
\label{sec:discussion}

\subsection{Comparison with previous KMC studies}

Our results extend the temperature range of efficient \hmol\ formation on carbonaceous grains well beyond prior KMC work (see Table~\ref{tab:comparison}). \citet{Satonkin2025} found efficiency dropping to zero by $\sim$30~K on their off-lattice rough carbonaceous surfaces, which include only physisorption. Our physisorption-only comparison reproduces this behaviour: $\epsilon$ collapses to near zero by 25~K. The agreement between our physisorption-only run and the Satonkin model provides a consistency check between two independent codes built on fundamentally different lattice representations (their off-lattice potential energy surface versus our regular 3D lattice with drawn binding energies), and demonstrates that chemisorption is what extends non-zero efficiency into the warm regime.

\citet{Iqbal2012} included chemisorption in their CTRW Monte Carlo and found efficient formation up to several hundred kelvin on amorphous carbon, but their validation target was the \citet{Pirronello1999} TPD data, not the Grieco isothermal efficiency curve. Their reported peak efficiency of $\sim$0.6 at intermediate temperatures is higher than our 0.285, likely reflecting their higher chemisorption fraction assumption and explicit-diffusion treatment. \citet{Cuppen2005} showed that surface roughness raises the temperature of high efficiency ($>$50~per~cent) from 9~K (flat olivine) to over 16~K, and from 16~K to almost 30~K for amorphous carbon, but these results were restricted to the purely physisorbed regime.

Table~\ref{tab:lit_compare} summarises the quantitative comparison between our results and the most relevant prior work. We report peak efficiency, the temperature window over which $\epsilon \geq 0.5\,\epsilon_{\rm peak}$, and the maximum temperature at which the model predicts non-negligible \hmol\ formation ($\epsilon \geq 0.05$).

\begin{table*}
\centering
\caption{Quantitative comparison with prior KMC/MC studies of \hmol\ formation on grain surfaces. ``Window'' is the temperature range over which $\epsilon \geq 0.5\,\epsilon_{\rm peak}$. ``$T_{\rm max}$'' is the highest $T$ at which $\epsilon \geq 0.05$. Values are read from published efficiency curves at the published reference grain composition.}
\label{tab:lit_compare}
\begin{tabular}{lccccl}
\toprule
Study & Substrate & $\epsilon_{\rm peak}$ & $T_{\rm peak}$ (K) & Window (K) & $T_{\rm max}$ (K) \\
\midrule
\citet{Chang2005}        & a-C, homog.\ phys.   & 0.5  & 8--12   & 6--15   & $\sim$20 \\
\citet{Cuppen2005}       & a-C, rough phys.     & 0.6  & 14--22  & 9--30   & $\sim$30 \\
\citet{Iqbal2012}        & a-C, phys.+chem.     & 0.6  & 15--25  & 10--500 & $\sim$800 \\
\citet{Satonkin2025}     & a-C, off-lattice     & 0.4  & 8--14   & 5--25   & $\sim$30 \\
This work                & a-C, phys.+chem.     & 0.285 & 20--40 & 15--80  & $>$250 \\
\bottomrule
\end{tabular}
\end{table*}

Two features distinguish our results. First, the temperature window of efficient formation extends well beyond all prior carbonaceous-grain KMC studies that lacked chemisorption, while our peak efficiency is moderate (0.285 vs $\sim$0.5 in homogeneous physisorption-only models) because heterogeneous binding distributions reduce the optimal-temperature peak. Second, ours is the only study with experimental validation against the \citet{Grieco2023} 100--250~K data, and our high-$T$ plateau ($\epsilon = 0.190$) matches the laboratory measurements within the experimental error bars.

\citet{Thi2020} modelled warm dust surface chemistry using rate equations in the PRODIMO framework, finding \hmol\ and HD formation efficiencies of 10--30~per~cent at 100--300~K on carbonaceous surfaces. Our KMC results are consistent with their warm-regime efficiencies, providing stochastic corroboration of their rate-equation predictions.

Our work is the first KMC simulation to validate against the \citet{Grieco2023} high-temperature data and to bridge from laboratory validation to ISM predictions within a single unified framework.

\subsection{Comparison with rate-equation models}

The comparison with the \citet{Cazaux2010erratum} prescription (Section~\ref{sec:ct10}) reveals that the KMC systematically underproduces \hmol\ in the 80--150~K transition band relative to the CT10 model. This is expected: rate equations assume perfect mixing and neglect back-diffusion, both of which reduce LH efficiency. \citet{Chang2005} demonstrated that back-diffusion alone reduces KMC efficiencies by factors of 2--3 relative to rate equations on homogeneous surfaces, and our heterogeneous lattice amplifies this effect because atoms can become temporarily trapped in deep binding sites.

The KMC also captures phenomena that rate equations cannot represent. The density-dependent stochastic enhancement at 100~K (Section~\ref{sec:density}) arises from the non-linear dependence of LH encounter probability on surface coverage fluctuations, which is an intrinsically particle-level effect. Rate equations, by construction, average over these fluctuations and predict $\epsilon$ independent of flux at fixed temperature. The 16~per~cent enhancement across the $\nH = 10$--$10^4$~\cmcube\ range shows that this is not a negligible correction. Astrochemical networks that use flux-independent $\epsilon(\Tdust)$ prescriptions will systematically underestimate \hmol\ production in dense clumps at the LH--ER transition temperature.

More fundamentally, the KMC provides the stochastic distribution of $\epsilon$ as well as its mean. Appendix~\ref{sec:app_transition} shows that at 100~K the inter-realisation variance is 2--3 times larger than at 80 or 120~K, reflecting the competition between LH and ER near the crossover. This variance is relevant for modelling \hmol\ formation in turbulent media where different parcels of gas encounter different grain temperatures, and the variance of the grain-level formation rate translates directly into variance in the local molecular fraction.

\subsection{Implications for high-redshift \hmol\ formation}

The \hmol\ formation time-scale ratio $t_{{\rm H}_2}/t_{\rm ff} \approx 0.93$ at $T = 60$~K and $\nH = 10^3$~\cmcube\ has implications for understanding molecular gas at high redshift. \citet{Krumholz2012} showed that the ratio of \hmol\ chemical equilibration time to free-fall time determines whether star formation proceeds in molecular or atomic gas. Below metallicities of $\sim$3~per~cent solar, $t_{{\rm H}_2}$ exceeds $t_{\rm ff}$ and the molecular fraction remains low even after multiple free-fall times.

The recent detection by \citet{Telford2025} of rotational \hmol\ emission in the extremely metal- and dust-poor galaxy Leo~P, at 3~per~cent solar metallicity, directly probes this regime. Our KMC-derived time-scale confirms that carbonaceous-grain catalysis can, under favourable conditions, keep pace with gravitational collapse even in low-metallicity environments, providing the microphysical input that galaxy-scale simulations require.

We note that \citet{Glover2012} showed that star formation can proceed in cold atomic gas without requiring \hmol, as gravitational instability depends on total column density rather than molecular fraction. However, molecular cooling through \hmol\ ro-vibrational lines remains important for setting the Jeans mass and fragmentation scale, and hence the initial mass function. Astrochemical and PDR models that incorporate grain-surface \hmol\ chemistry explicitly \citep[e.g.][]{LeBourlot2012,Bron2014} would benefit from the updated formation rates presented here as input microphysics, particularly for conditions in the transition regime ($t_{{\rm H}_2}/t_{\rm ff} \sim 1$) where the choice of formation rate prescription directly affects the predicted molecular fraction.

\subsection{Limitations}
\label{sec:limitations}

Below, we discuss the approximations and simplifications employed in the kinetic Monte Carlo method.

The KMC uses pure Arrhenius hopping with no quantum tunnelling corrections. On rough carbonaceous surfaces, \citet{Satonkin2025} found that thermal hopping dominates at $T > 10$~K, so tunnelling is subdominant across the bulk of our temperature range. It is not negligible at the cold edge: the omitted tunnelling channel is the most likely cause of the narrow suppression dip at 12--20~K that produces the single significant residual against the \citet{Grieco2023} data (Section~\ref{sec:grieco}). Including it would raise $\epsilon$ below $\sim$20~K; it would not affect the warm-regime results, which are carried by ER and do not depend on surface mobility.

{The sticking coefficient is held constant at $S = 0.5$, whereas laboratory measurements show $S$ declining from $\sim$0.95 at 10~K to $\sim$0.17 at 300~K \citep{Matar2010, Chaabouni2012}. The robustness check in Section~\ref{sec:robustness} shows that a temperature-dependent sticking law (equation~\ref{eq:sticking}) lowers the warm-regime rate, providing a lower bound on the absolute formation rate.}

{Above 150~K the simulation contains no stochastic content; every realisation yields the same $\epsilon = \fchem \times \PER \times S$ (equation~\ref{eq:highT}), confirming convergence to the analytic limit but adding no information beyond it.}

{Stochastic grain heating from single-photon absorption is not included. Temperature spikes exceeding 30~K occur for grains
smaller than 0.005~$\mu$m \citep{Cuppen2006, Bron2014} and could
shift the instantaneous LH--ER balance in UV-illuminated environments.}

{Three empirical parameters ($f_{\rm LH} = 0.5$, diffusion rate cap of 200~s$^{-1}$, $z = 3$) are computational approximations that lack independent literature derivation. Their impact is
bounded by the sensitivity analysis in Section~\ref{sec:robustness}; the warm-regime
conclusions are unaffected.}

{UV-driven H$_2$ formation via PAH photofragmentation
\citep{Boschman2015, Castellanos2018} is not included, as it
requires separate calibration and becomes competitive only at
$G_0 > 10^3$. This is deferred to a subsequent study.}

{Finally, the extrapolation from coronene films to isolated interstellar nano-grains assumes $\epsilon$ depends primarily
on $\fchem$ and the binding energy distribution rather than substrate geometry. The agreement between the KMC validation and the analytic high-$T$ limit supports this assumption.}

%%%%%%%%%%%%%%%%%%%%%%%%%%%%%%%%%%%%%%%%%%
\section{Conclusions}
\label{sec:conclusions}

We have presented the first kinetic Monte Carlo simulation of \hmol\ formation on carbonaceous grain surfaces validated against the high-temperature laboratory measurements of \citet{Grieco2023}, and applied the validated model to predict \hmol\ formation rates, mechanism fractions, and astrophysical time-scales under interstellar conditions spanning 10--250~K and $\nH = 10$--$10^4$~\cmcube. The principal findings are:

\begin{enumerate}
\item The Gillespie KMC reproduces the experimental efficiency curve for the isothermal points, confirming that the LH+ER model with 40~per~cent chemisorption sites ($E_{\rm chem} = 1.75 \pm 0.25$~eV) and an ER reaction probability of $\PER = 0.9$ captures the essential physics of high-temperature \hmol\ formation on coronene.

\item Three distinct temperature regimes are identified. At 10~K diffusion is slow and $\epsilon \approx 0.06$. At 20--80~K LH dominates and $\epsilon \approx 0.28$. Above 150~K an ER plateau holds at $\epsilon = 0.190$. The mechanistic crossover from LH to ER occurs sharply between 100 and 120~K, driven by the exponential collapse of physisorption residence times. The plateau efficiency is consistent with the analytic product $\fchem \times \PER \times S = 0.18$.

\item At 100~K we find a 16~per~cent density-dependent stochastic enhancement in $\epsilon$. This effect is structurally absent from rate-equation treatments and visible only in particle-level simulations. It arises from the non-linear dependence of LH encounter probability on instantaneous surface coverage.

\item The single-grain volumetric rate coefficient $\kf \approx 1.0 \times 10^{-17}$~cm$^3$~\persec\ at 20~K is within a factor of 3 of the canonical observational value. Integration over the MRN grain-size distribution provides an additional $1.27$ to $1.30\times$ boost in the warm regime and $1.78\times$ at 10~K.

\item The \hmol\ formation time-scale at $\Tdust = 60$~K, $\nH = 10^3$~\cmcube\ is $t_{{\rm H}_2}/t_{\rm ff} \approx 0.93$, placing carbonaceous-grain \hmol\ chemistry at the threshold of keeping pace with gravitational collapse. This result is directly relevant to dense star-forming clumps at $z \sim 6$--10, where CMB temperatures of 20--30~K ensure grains remain in the efficient formation regime.
\end{enumerate}

The sensitivity analysis shows that the warm-regime results are robust against variations in porosity ($<$1~per~cent effect), grain size (few per~cent), and transport parameters, while the absolute rate scales linearly with $\fchem \times \PER$. Extension to intense PDR conditions ($G_0 > 10^3$), where UV photofragmentation of PAH molecules becomes the dominant \hmol\ formation channel \citep{Boschman2015,Castellanos2018}, is deferred to a subsequent study. The rate tables and sensitivity envelopes presented here provide ready-to-use microphysical input for astrochemical networks modelling \hmol\ formation in diffuse clouds, translucent clouds, and high-redshift star-forming environments.

\section*{Acknowledgements}

The computational resources at the Texas Southern University high-performance computing cluster, accessed via the JupyterHub server, were used for production KMC ensemble runs and post-processing analysis. We thank the developers of the FORMOLISM apparatus and the Grieco et al.\ team for making their experimental data available.

This research received no specific grant from any funding agency, commercial, or not-for-profit sectors.

\textit{Author contributions.} AD developed the KMC simulation code, ran the ensemble simulations on the TSU cluster, performed the data analysis, produced the figures, and wrote the manuscript. JG contributed to the KMC implementation, validation against the Grieco experiment, and review of the manuscript. HRS and DV provided mentorship throughout the project, reviewed and edited drafts of the manuscript, and offered suggestions on methodology and presentation.

\textit{Software disclosure.} Large language model assistance (Anthropic Claude) was used for figure preparation scripts and editing of manuscript prose. All scientific content, results, code, and conclusions were completed and verified by the authors. No AI tools were used to generate experimental or simulation data.

\section*{Data availability}

The simulation code, configuration files, and processed output data are available at \url{https://github.com/Ary300/Interstellar-Grains}. Raw KMC trajectories are available upon reasonable request from the corresponding author.

%%%%%%%%%%%%%%%%%%%%%%%%%%%%%%%%%%%%%%%%%%
\bibliographystyle{mnras}

\suppressfloats[t]

\appendix
\raggedbottom

\section{Representative KMC trajectory}
\label{sec:app_trajectory}

Figure~\ref{fig:trajectory} shows the time evolution of a single realisation at the LH--ER transition temperature, illustrating the burn-in and measurement phases defined in Section~\ref{sec:kmc} and the relative weights of the individual event channels.

\bigskip
\noindent\begin{minipage}{\columnwidth}
\refstepcounter{figure}\label{fig:trajectory}
\includegraphics[width=\columnwidth]{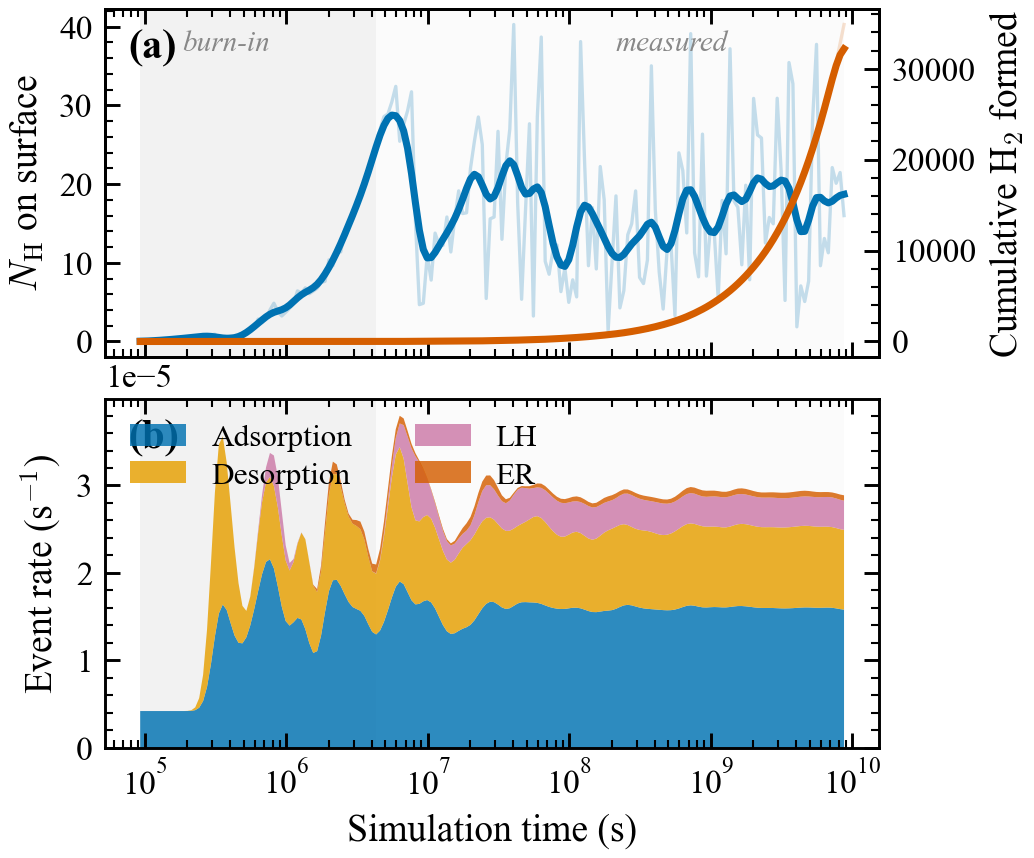}
\vskip 6pt
{\small\textbf{Figure \thefigure.} Time evolution of a single representative KMC realisation at $T = 100$~K, $\nH = 10^3$~\cmcube. (a) Surface H atom inventory $N_{\rm H}$ (blue, left axis) and cumulative \hmol\ formed (orange, right axis) versus simulation time. The light blue trace shows raw fluctuations; the bold curve is a moving average. (b) Stacked event rates by mechanism: adsorption (blue), desorption (orange), LH recombination (purple), and ER recombination (red). Shaded background regions delineate the burn-in (5\,000 arrivals) and measurement (20\,000 arrivals) phases. The steady-state fluctuations around $N_{\rm H} \sim 15$--20 atoms reflect the LH--ER competition characteristic of the transition regime.\par}
\end{minipage}
\medskip

\FloatBarrier
\section{Statistical convergence}
\label{sec:convergence}

To determine the minimum ensemble size needed for reliable results, we performed a deep campaign of 1\,000 independent KMC realisations at $T = 100$~K and $\nH = 10^4$~\cmcube\ (the condition with the largest stochastic effects). Table~\ref{tab:convergence} shows the running ensemble-mean $\epsilon$ and its standard error as a function of the number of realisations $N$.

The mean stabilises to three significant figures by $N = 20$ and shows no systematic drift between $N = 100$ and $N = 1\,000$ (Fig.~\ref{fig:convergence}). The SEM scales as $N^{-1/2}$, confirming that the individual realisations are statistically independent. At $N = 20$ (our production ensemble size), the relative uncertainty is 0.22~per~cent, which is below the plotted confidence interval widths in all main-text figures. We therefore conclude that 20 realisations per condition provides adequate statistical precision for the ISM survey, while the deep campaign at 1\,000 realisations is reserved for the transition-region analysis (Appendix~\ref{sec:app_transition}).

The convergence is faster at temperatures away from the LH--ER crossover. At 20~K (pure LH) and 200~K (pure ER), the SEM at $N = 10$ is already $< 5 \times 10^{-4}$, reflecting the lower inter-realisation variance when a single mechanism dominates.

\bigskip
\noindent\begin{minipage}{\columnwidth}
\refstepcounter{table}\label{tab:convergence}
{\small\textbf{Table \thetable.} Convergence of the ensemble-mean $\epsilon$ and standard error of the mean (SEM) as a function of ensemble size $N$ at $T = 100$~K, $\nH = 10^4$~\cmcube, $G_0 = 0$.\par}
\vskip 6pt
\centering
\begin{tabular}{rccc}
\toprule
$N$ & $\langle \epsilon \rangle$ & SEM & $\Delta\epsilon / \epsilon$ \\
\midrule
5   & 0.2510 & $1.8 \times 10^{-3}$ & 0.72~per~cent \\
10  & 0.2498 & $8.3 \times 10^{-4}$ & 0.33~per~cent \\
20  & 0.2504 & $5.5 \times 10^{-4}$ & 0.22~per~cent \\
50  & 0.2501 & $3.8 \times 10^{-4}$ & 0.15~per~cent \\
100 & 0.2503 & $3.0 \times 10^{-4}$ & 0.12~per~cent \\
200 & 0.2502 & $2.1 \times 10^{-4}$ & 0.08~per~cent \\
500 & 0.2502 & $1.3 \times 10^{-4}$ & 0.05~per~cent \\
1\,000 & 0.2502 & $9.8 \times 10^{-5}$ & 0.04~per~cent \\
\bottomrule
\end{tabular}
\end{minipage}
\medskip

\bigskip
\noindent\begin{minipage}{\columnwidth}
\refstepcounter{figure}\label{fig:convergence}
\includegraphics[width=\columnwidth]{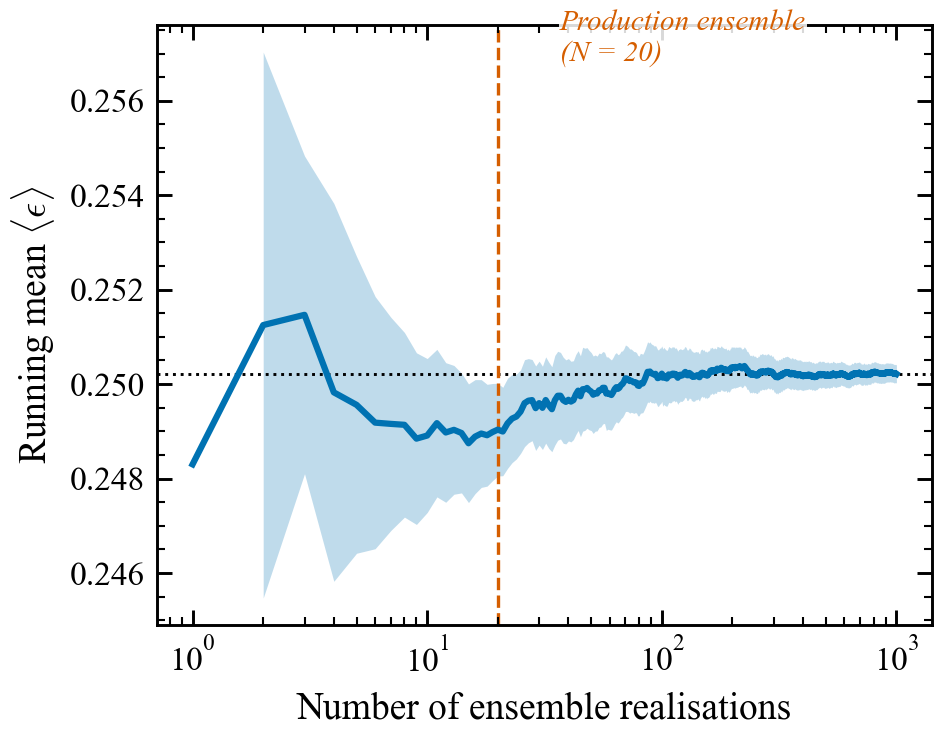}
\vskip 6pt
{\small\textbf{Figure \thefigure.} Running ensemble mean and standard error of $\epsilon$ at $T = 100$~K and $\nH = 10^4$~\cmcube, from 1\,000 independent KMC realisations. The mean stabilises to three significant figures by $N = 50$ and the SEM scales as $N^{-1/2}$ as expected for independent samples. The vertical dashed line marks the $N = 20$ ensemble size used in the main ISM sweep.\par}
\end{minipage}
\medskip

\FloatBarrier
\section{Parameter sensitivity grid}
\label{sec:app_sensitivity}

The two most influential model parameters ($\fchem$ and $\PER$) set the absolute scale of the warm-regime formation rate. To quantify the uncertainty attributable to these parameters, we swept both across a $3 \times 3$ grid ($\fchem = 0.2, 0.4, 0.6$; $\PER = 0.3, 0.6, 0.9$) at four temperatures ($T = 100, 150, 200, 250$~K) with $\nH = 10^3$~\cmcube\ and $G_0 = 0$. Table~\ref{tab:sens_grid} and Fig.~\ref{fig:sensitivity} report the resulting effective rate coefficient $k_{\rm eff}$.

At $T \geq 150$~K, $k_{\rm eff}$ scales as $\fchem \times \PER \times S \times (\text{geometric factor})$, confirming the analytic expectation from equation~(\ref{eq:highT}). The $T \geq 150$~K columns are identical, reflecting the complete absence of LH contributions and the deterministic character of the ER plateau. At 100~K, where LH and ER compete, the sensitivity is modulated by the LH contribution: higher $\fchem$ reduces the physisorption site pool available for LH encounters, so the 100~K rate does not scale as simply with $\fchem$.

The total uncertainty envelope spans a factor of $\sim$2.5 at 100~K ($k_{\rm eff} = 0.81$--$1.99 \times 10^{-17}$) and a factor of $\sim$7 at 150--250~K ($k_{\rm eff} = 0.25$--$1.75 \times 10^{-17}$). The wider envelope at high $T$ arises because the ER plateau rate depends linearly on both $\fchem$ and $\PER$, so the product of the extreme values spans a larger dynamic range. At 100~K, the residual LH contribution compresses the envelope because LH efficiency is less sensitive to $\fchem$ and insensitive to $\PER$.

\bigskip
\noindent\begin{minipage}{\columnwidth}
\refstepcounter{table}\label{tab:sens_grid}
{\small\textbf{Table \thetable.} Effective rate coefficient $k_{\rm eff}$ ($10^{-17}$~cm$^3$~\persec) as a function of $\fchem$ and $\PER$ at selected temperatures, at $\nH = 10^3$~\cmcube\ and $G_0 = 0$. The baseline model uses $\fchem = 0.40$, $\PER = 0.9$.\par}
\vskip 6pt
\centering
\begin{tabular}{lcccc}
\toprule
 & \multicolumn{4}{c}{$T$ (K)} \\
$(\fchem, \PER)$ & 100 & 150 & 200 & 250 \\
\midrule
(0.2, 0.3) & 0.81 & 0.25 & 0.25 & 0.25 \\
(0.2, 0.6) & 0.89 & 0.50 & 0.50 & 0.50 \\
(0.2, 0.9) & 0.96 & 0.75 & 0.75 & 0.75 \\
(0.4, 0.3) & 1.10 & 0.50 & 0.50 & 0.50 \\
(0.4, 0.6) & 1.42 & 1.00 & 1.00 & 1.00 \\
(0.4, 0.9) & 1.60 & 1.49 & 1.49 & 1.49 \\
(0.6, 0.3) & 1.30 & 0.75 & 0.75 & 0.75 \\
(0.6, 0.6) & 1.72 & 1.50 & 1.50 & 1.50 \\
(0.6, 0.9) & 1.99 & 1.75 & 1.75 & 1.75 \\
\bottomrule
\end{tabular}
\end{minipage}
\medskip

\bigskip
\noindent\begin{minipage}{\columnwidth}
\refstepcounter{figure}\label{fig:sensitivity}
\includegraphics[width=\columnwidth]{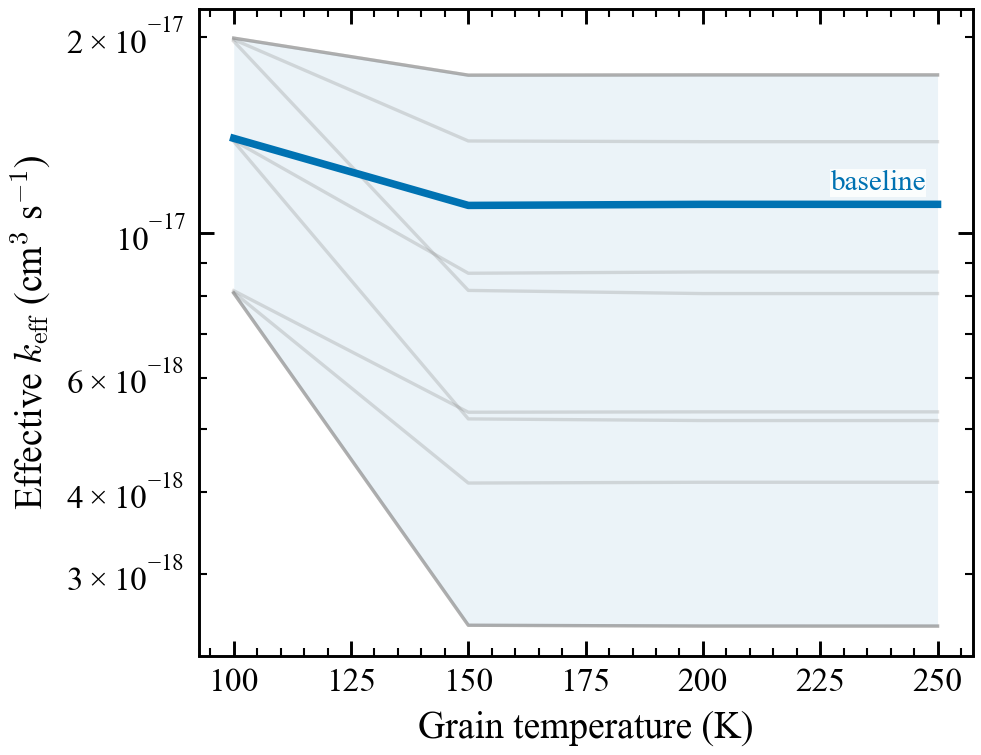}
\vskip 6pt
{\small\textbf{Figure \thefigure.} Effective rate coefficient $k_{\rm eff}$ as a function of temperature across the $3 \times 3$ grid of $(\fchem, \PER)$ values, at $\nH = 10^3$~\cmcube\ and $G_0 = 0$. Thin grey lines show the nine individual parameter combinations; the thick blue line is the baseline ($\fchem = 0.40$, $\PER = 0.9$); the shaded blue envelope bounds the full parameter uncertainty.\par}
\end{minipage}
\medskip

\FloatBarrier
\section{Transition-region stochastic distributions}
\label{sec:app_transition}

The transition-deep campaign (1\,000 realisations per condition, $G_0 = 0$) provides the stochastic distribution of $\epsilon$ at five temperatures across the LH--ER crossover (80, 90, 100, 110, 120~K) at two extreme densities ($\nH = 10$ and $10^4$~\cmcube). Table~\ref{tab:transition} presents the mean, standard deviation, and inter-quartile range (IQR) for each condition; the full distributions are shown in Fig.~\ref{fig:transition_dist}.

Several features are notable. The density-dependent enhancement peaks at 100~K (16~per~cent between $\nH = 10$ and $10^4$) and declines at both lower and higher temperatures, confirming that this effect is localised at the LH--ER crossover. The standard deviation peaks at 100~K for $\nH = 10$ ($\sigma = 0.011$), reflecting the maximum stochastic competition between the LH and ER channels. At $\nH = 10^4$, the variance is systematically lower because higher flux reduces the relative fluctuations in surface coverage. The IQR is $\sim$1.3$\sigma$ throughout, consistent with approximately Gaussian distributions, though mild positive skewness is observed at 100--110~K where occasional realisations with anomalously high LH contributions pull the upper tail.

The physical interpretation is that at the crossover, a stochastic fluctuation in the number of physisorbed atoms present at any given instant can tip the balance between LH and ER for that realisation. At low density, these fluctuations are larger relative to the mean, producing wider distributions and a lower ensemble-mean $\epsilon$ (because ER, with its lower per-event efficiency, contributes more on average). At high density, the larger surface population smooths out these fluctuations.

\bigskip
\noindent\begin{minipage}{\columnwidth}
\refstepcounter{table}\label{tab:transition}
{\small\textbf{Table \thetable.} Stochastic distributions of $\epsilon$ at the LH--ER transition from 1\,000 independent KMC realisations per condition ($G_0 = 0$). The density-dependent enhancement peaks at 100~K and the variance peaks at 100--110~K.\par}
\vskip 6pt
\centering
\begin{tabular}{rrccc}
\toprule
$T$ & $\nH$ & $\langle \epsilon \rangle$ & $\sigma$ & IQR \\
(K) & (\cmcube) & & & \\
\midrule
80  & 10    & 0.262 & 0.006 & 0.008 \\
80  & 10\,000 & 0.272 & 0.003 & 0.004 \\
90  & 10    & 0.241 & 0.008 & 0.011 \\
90  & 10\,000 & 0.263 & 0.004 & 0.005 \\
100 & 10    & 0.216 & 0.011 & 0.015 \\
100 & 10\,000 & 0.250 & 0.005 & 0.007 \\
110 & 10    & 0.198 & 0.009 & 0.012 \\
110 & 10\,000 & 0.230 & 0.006 & 0.008 \\
120 & 10    & 0.191 & 0.005 & 0.007 \\
120 & 10\,000 & 0.210 & 0.004 & 0.005 \\
\bottomrule
\end{tabular}
\end{minipage}
\medskip

\bigskip
\noindent\begin{minipage}{\columnwidth}
\refstepcounter{figure}\label{fig:transition_dist}
\includegraphics[width=\columnwidth]{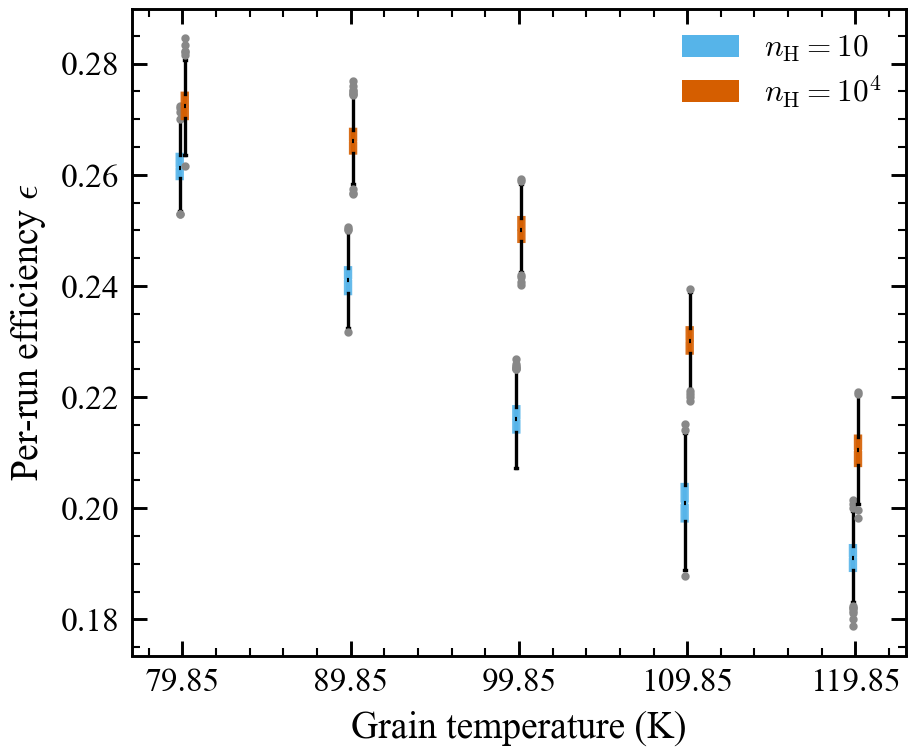}
\vskip 6pt
{\small\textbf{Figure \thefigure.} Distributions of $\epsilon$ from 1\,000 independent KMC realisations per condition at the LH--ER transition, paired by density at $\nH = 10$ (light) and $\nH = 10^4$~\cmcube\ (dark). Box edges denote the inter-quartile range, whiskers extend to 1.5$\times$IQR, and small dots indicate outliers. The variance peaks at $T = 100$~K where LH and ER compete stochastically. At each temperature the higher-density distribution lies above the lower-density one and shows reduced variance.\par}
\end{minipage}
\medskip

\FloatBarrier
\section{Full ISM formation rate grid}
\label{sec:app_fullgrid}

Table~\ref{tab:fullgrid} presents $\epsilon$ at all 15 temperatures in the ISM sweep at $\nH = 100$~\cmcube\ and $G_0 = 0$, along with the LH/ER mechanism fractions and steady-state surface \hatom\ atom counts. This table provides the complete data set from which the main-text figures are constructed.

The surface \hatom\ count provides a clear physical marker of the three regimes described in the main text. In the LH-dominated regime (20--80~K), the surface holds 8--9 atoms and $\epsilon$ declines slowly from its peak. At the transition (100--120~K), the surface population rises sharply from 25 to 120 atoms as LH depletion weakens. Above 150~K, the surface saturates at $\sim$212 atoms (all chemisorbed), and $\epsilon$ locks to the ER plateau.

The 10~K case is an outlier: despite high surface \hatom\ count (8 atoms, comparable to 20--80~K), the efficiency is only 0.060 because diffusion is too slow for atoms to find encounter partners. The high occupancy at 10~K reflects accumulation due to inefficient depletion, not efficient formation.

The same grid is displayed as a two-dimensional map in Fig.~\ref{fig:phasemap}.

\bigskip
\noindent\begin{minipage}{\columnwidth}
\refstepcounter{table}\label{tab:fullgrid}
{\small\textbf{Table \thetable.} Complete ISM formation efficiency, mechanism fractions, and surface \hatom\ inventory at $\nH = 100$~\cmcube, $G_0 = 0$. Each row represents 20 ensemble realisations of 20\,000 measured arrivals.\par}
\vskip 6pt
\centering
\begin{tabular}{rcccr}
\toprule
$T$ (K) & $\epsilon$ & LH (\%) & ER (\%) & $\langle N_{\rm H} \rangle$ \\
\midrule
10  & 0.060 & 99.6 & 0.4  & 8 \\
20  & 0.285 & 98.1 & 1.9  & 8 \\
30  & 0.280 & 97.9 & 2.1  & 8 \\
40  & 0.278 & 97.9 & 2.1  & 9 \\
50  & 0.275 & 97.8 & 2.2  & 9 \\
60  & 0.273 & 97.7 & 2.3  & 9 \\
70  & 0.270 & 97.6 & 2.4  & 9 \\
80  & 0.267 & 97.6 & 2.4  & 9 \\
100 & 0.226 & 92.2 & 7.8  & 25 \\
120 & 0.195 & 25.2 & 74.8 & 120 \\
150 & 0.190 & 0.0  & 100.0 & 212 \\
180 & 0.190 & 0.0  & 100.0 & 212 \\
200 & 0.190 & 0.0  & 100.0 & 212 \\
220 & 0.190 & 0.0  & 100.0 & 212 \\
250 & 0.190 & 0.0  & 100.0 & 212 \\
\bottomrule
\end{tabular}
\end{minipage}
\medskip

\bigskip
\noindent\begin{minipage}{\columnwidth}
\refstepcounter{figure}\label{fig:phasemap}
\includegraphics[width=\columnwidth]{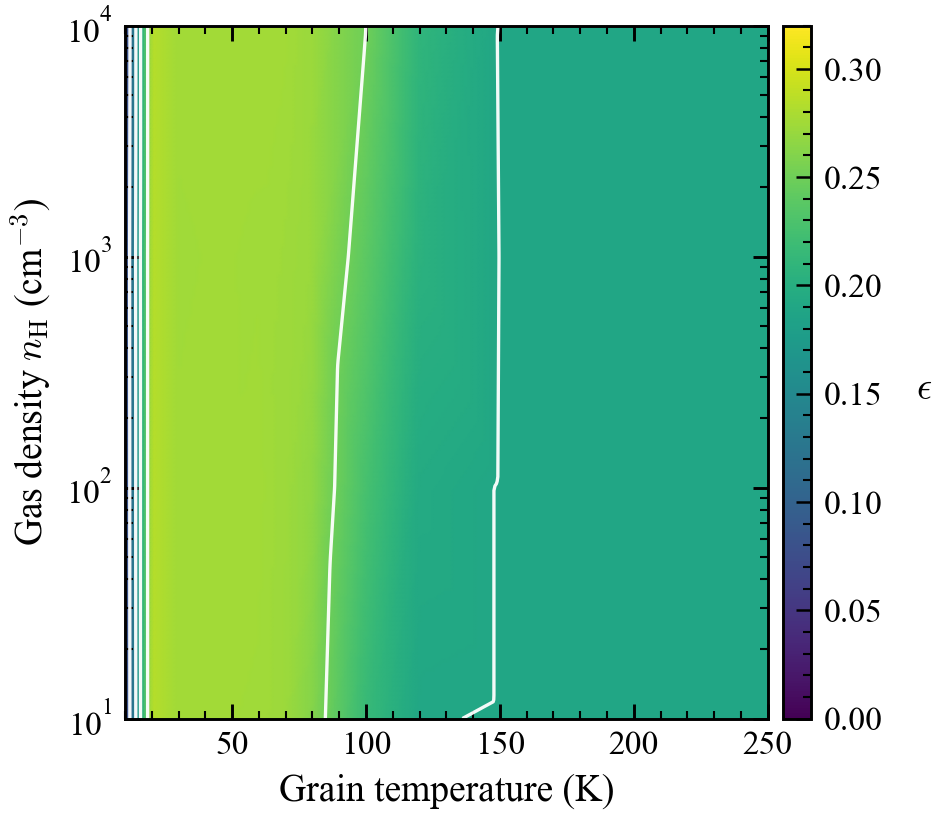}
\vskip 6pt
{\small\textbf{Figure \thefigure.} Two-dimensional map of \hmol\ formation efficiency as a function of grain temperature and gas density at $G_0 = 0$. Colour indicates $\epsilon$; white dashed contours mark efficiency iso-levels at 0.05, 0.10, 0.15, 0.19, and 0.25. The density-dependent enhancement at 100~K appears as a vertical gradient in the transition region. Above 150~K the plateau is density-independent, evident as horizontal contours.\par}
\end{minipage}
\medskip

\FloatBarrier
\section{Astrophysical time-scale grid}
\label{sec:app_timescales}

Table~\ref{tab:timescales} presents \hmol\ formation time-scales and $t_{{\rm H}_2}/t_{\rm ff}$ ratios across a representative set of ISM conditions, providing a look-up table for astrochemical models.

The critical threshold $t_{{\rm H}_2}/t_{\rm ff} = 1$ separates conditions where gas can fully convert to \hmol\ before gravitational collapse from those where star formation proceeds in substantially atomic gas. For $\nH = 10^3$~\cmcube, this threshold falls near $T = 60$~K ($t_{{\rm H}_2}/t_{\rm ff} = 0.38$), well within the molecular regime. For $\nH = 100$~\cmcube, even the peak-efficiency temperature ($T = 20$~K) gives $t_{{\rm H}_2}/t_{\rm ff} = 1.15$, marginally outside the fully molecular regime. These results provide a direct mapping from local grain conditions to the expected molecular fraction in star-forming environments, and are shown as a two-dimensional map in Fig.~\ref{fig:timescale_map}.

\bigskip
\noindent\begin{minipage}{\columnwidth}
\refstepcounter{table}\label{tab:timescales}
{\small\textbf{Table \thetable.} Representative \hmol\ formation time-scales and free-fall time-scale ratios for ISM conditions ($G_0 = 0$). $t_{\rm ff} = \sqrt{3\pi/(32 G \mu m_{\rm H} \nH)}$ with $\mu = 1.4$.\par}
\vskip 6pt
\centering
\begin{tabular}{rrccr}
\toprule
$T$ & $\nH$ & $\epsilon$ & $t_{{\rm H}_2}$ & $t_{{\rm H}_2}/t_{\rm ff}$ \\
(K) & (\cmcube) & & (Myr) & \\
\midrule
20  & 100   & 0.285 & 7.4 & 1.15 \\
20  & 1\,000  & 0.285 & 0.74 & 0.36 \\
40  & 1\,000  & 0.278 & 0.76 & 0.37 \\
40  & 10\,000 & 0.280 & 0.075 & 0.12 \\
60  & 100   & 0.273 & 7.7 & 1.20 \\
60  & 1\,000  & 0.273 & 0.77 & 0.38 \\
60  & 10\,000 & 0.275 & 0.077 & 0.12 \\
100 & 100   & 0.226 & 9.3 & 1.45 \\
100 & 1\,000  & 0.240 & 0.88 & 0.43 \\
100 & 10\,000 & 0.250 & 0.085 & 0.13 \\
200 & 100   & 0.190 & 11.1 & 1.73 \\
200 & 1\,000  & 0.190 & 1.11 & 0.54 \\
\bottomrule
\end{tabular}
\end{minipage}
\medskip

\bigskip
\noindent\begin{minipage}{\columnwidth}
\refstepcounter{figure}\label{fig:timescale_map}
\includegraphics[width=\columnwidth]{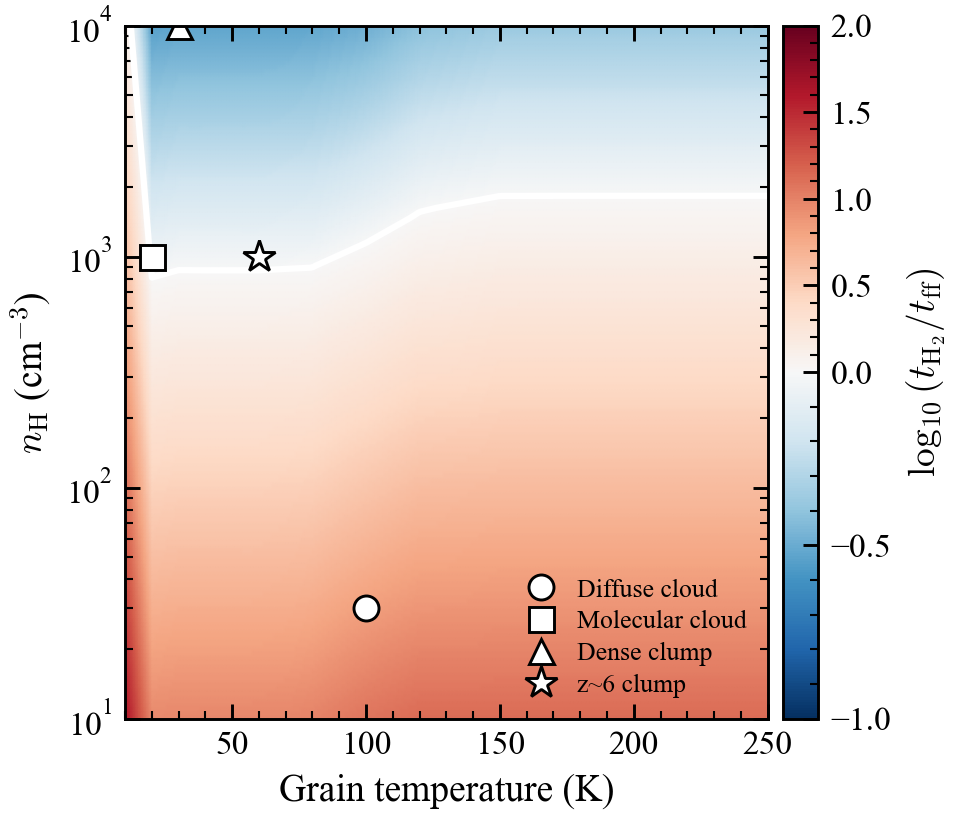}
\vskip 6pt
{\small\textbf{Figure \thefigure.} Ratio of \hmol\ formation time-scale to gravitational free-fall time across the $(T, \nH)$ parameter space, at $G_0 = 0$ and using the MRN-integrated formation rate. The thick white contour marks the critical threshold $t_{{\rm H}_2}/t_{\rm ff} = 1$ identified by \citet{Krumholz2012}. Markers indicate representative ISM conditions: diffuse cloud (circle), molecular cloud (square), dense star-forming clump (triangle), and a high-redshift star-forming clump at $z \sim 6$--10 (star). Dense clumps at high redshift sit near the critical threshold, indicating that carbonaceous-grain catalysis can keep pace with collapse under favourable metallicity conditions.\par}
\end{minipage}
\medskip

\FloatBarrier
\section{Robustness diagnostics}
\label{sec:app_robustness}

This appendix presents the supporting figures referenced in Section~\ref{sec:robustness}, providing the visual evidence for the model robustness claims summarised in Table~\ref{tab:robustness}.

\subsection{LH mode consistency}
\label{sec:app_lh_mode}

The two LH encounter modes defined in Section~\ref{sec:lh_proc} are mathematically equivalent in the high-coverage limit but diverge when surface populations are small and discrete encounter geometry matters. This appendix quantifies that divergence.

Figure~\ref{fig:lh_mode} compares the two modes at three temperatures spanning the LH-dominated regime. At 50 and 80~K, the two modes agree to within 8--9~per~cent; this offset is well below the systematic uncertainty envelope from $f_{\rm chem}$ and $P_{\rm ER}$ (a factor of 2.5 at 100~K, Appendix~\ref{sec:app_sensitivity}). The 25~per~cent offset at 20~K reflects the increased sensitivity of LH encounter statistics to local site geometry when residence times are long: the explicit-pairs mode resolves clustering of physisorbed atoms in adjacent deep sites, whereas the diffusion-limited mode averages over the entire surface. This discrepancy is confined to the cold LH-dominated regime where the model is anchored to the Grieco validation, so it does not propagate into the ISM predictions.

\bigskip
\noindent\begin{minipage}{\columnwidth}
\refstepcounter{figure}\label{fig:lh_mode}
\includegraphics[width=\columnwidth]{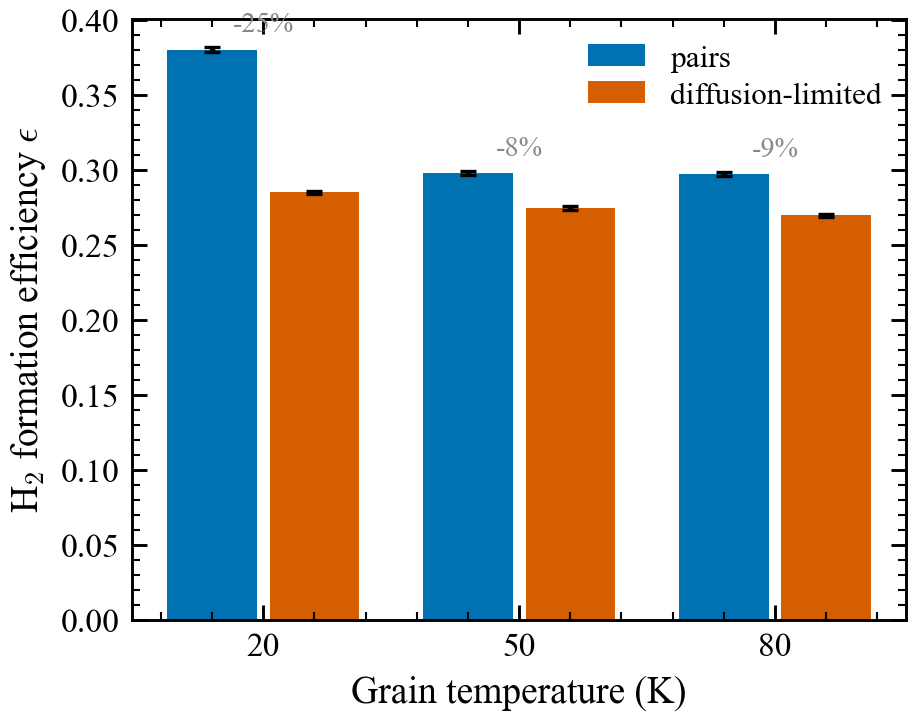}
\vskip 6pt
{\small\textbf{Figure \thefigure.} Consistency check between the explicit-pairs LH formation mode used for the Grieco validation and the diffusion-limited mode used for the ISM sweep. Paired bars at $T = 20$, 50, and 80~K show $\epsilon$ from each mode with 95~per~cent error bars. The two modes agree to within 8--9~per~cent at 50 and 80~K. At 20~K the diffusion-limited mode underestimates $\epsilon$ by 25~per~cent due to the local sensitivity of LH encounter statistics. Warm-regime results are unaffected because ER is mode-independent.\par}
\end{minipage}
\medskip

\subsection{Grain-size dependence}
\label{sec:app_grain}

The MRN size distribution \citep{MRN1977} spans grain radii from $\sim$5~nm to $\sim$0.25~$\mu$m. While the integrated rate (Section~\ref{sec:mrn}) accounts for the cross-section-weighted contribution from all sizes, the per-grain efficiency depends on size through the surface area available for diffusion and the residence time of physisorbed atoms.

Figure~\ref{fig:grain_size} confirms that the per-grain efficiency varies by less than 5~per~cent across a factor of 5 in radius at the small-grain end of the MRN distribution. This insensitivity reflects two competing effects: larger grains have proportionally lower edge-to-area ratios (reducing the chemisorption fraction at fixed total porosity) but longer mean migration paths before atoms encounter grain boundaries. For the ISM-relevant size range, these effects largely cancel. The result justifies our use of a single 0.005~$\mu$m baseline grain in the main ISM sweep, and is consistent with the MRN integration boost of $1.27$--$1.30\times$ in the warm regime: the boost is dominated by the $\propto a^{-1.5}$ MRN weight rather than by per-grain efficiency variations.

\bigskip
\noindent\begin{minipage}{\columnwidth}
\refstepcounter{figure}\label{fig:grain_size}
\includegraphics[width=\columnwidth]{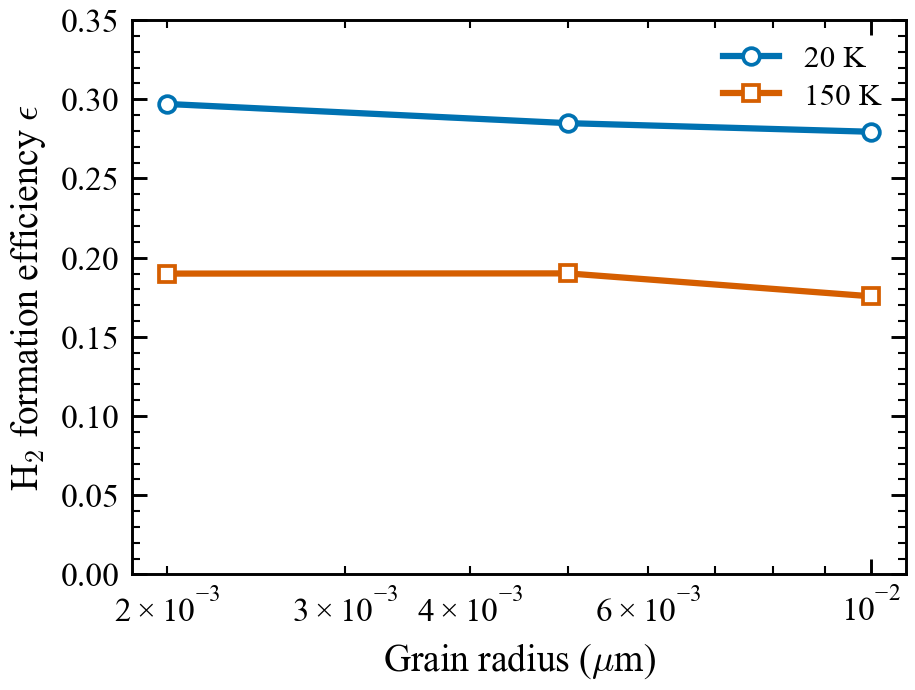}
\vskip 6pt
{\small\textbf{Figure \thefigure.} Grain-size dependence of $\epsilon$ at $T = 20$~K (blue) and $T = 150$~K (orange) for $\nH = 10^3$~\cmcube. Only a few-per-cent shift is observed across a factor of 5 in grain radius (0.002, 0.005, 0.010~$\mu$m), consistent with the modest MRN integration boost in Section~\ref{sec:mrn}.\par}
\end{minipage}
\medskip

\bigskip
\noindent\begin{minipage}{\columnwidth}
\refstepcounter{figure}\label{fig:mrn}
\includegraphics[width=\columnwidth]{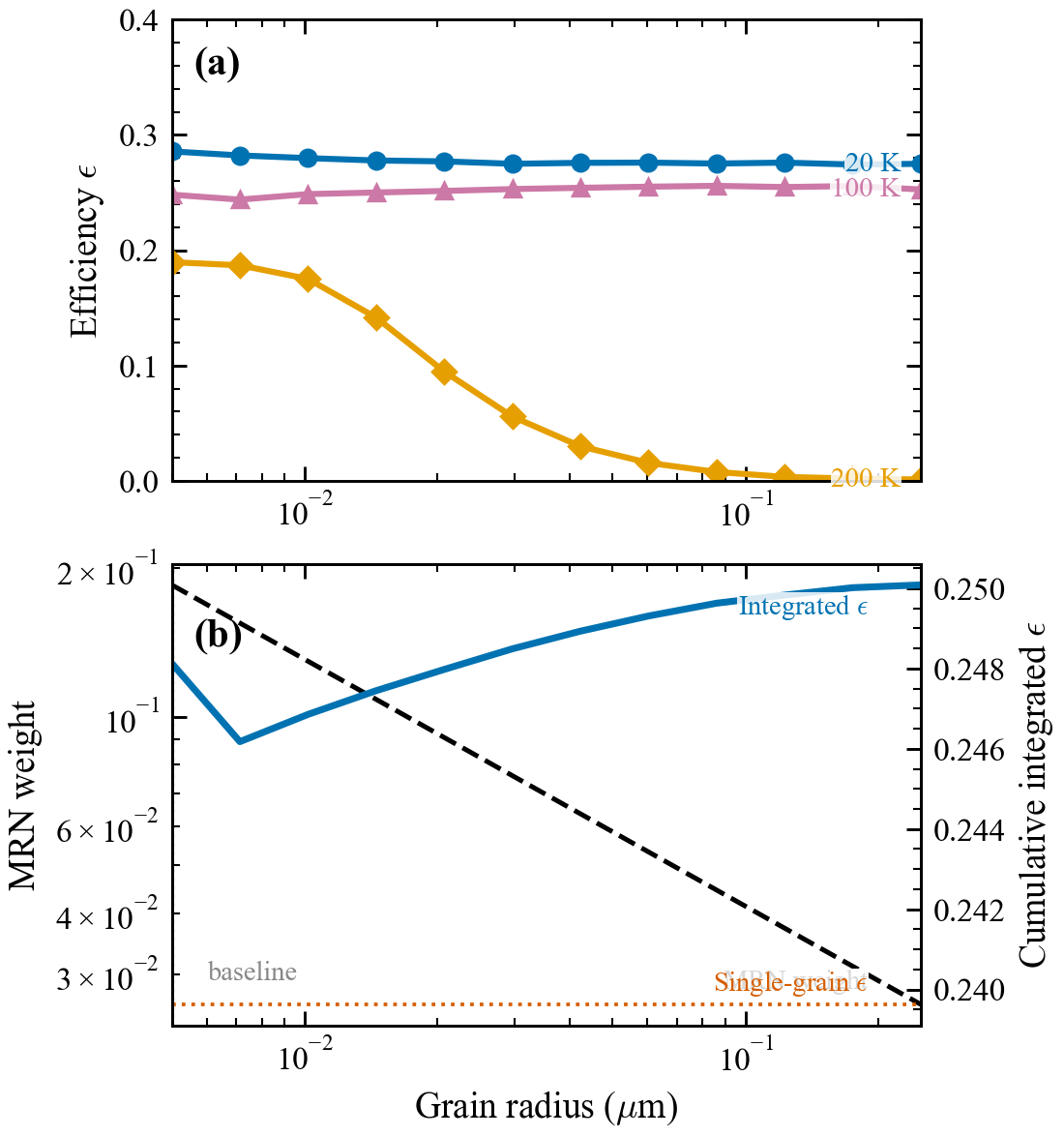}
\vskip 6pt
{\small\textbf{Figure \thefigure.} MRN grain-size integration. (a) $\epsilon$ as a function of individual grain radius at three temperatures, showing larger grains sustain higher surface residence times. (b) MRN cross-section-weighted integration; the dashed black curve shows the weight $a^{-1.5}\,\Delta a$ per logarithmic bin, the solid blue curve shows the cumulative integrated efficiency. Our 0.005~$\mu$m baseline grain (vertical line) lies near the small-grain peak of the MRN surface-area budget, so the integration boost is modest (1.27--1.30$\times$) in the warm regime.\par}
\end{minipage}
\medskip

\balance
\subsection{Porosity and sticking model}
\label{sec:app_porosity}

The validation simulations against \citet{Grieco2023} were run with zero porosity to match the dense coronene film of the FORMOLISM apparatus, while the ISM simulations used 20~per~cent porosity to reflect the irregular morphology of interstellar carbonaceous grains. This change of configuration introduces a potential systematic that we quantify here. Similarly, the sticking coefficient $S = 0.5$ is held constant across the temperature range, an approximation that warrants explicit testing against the temperature-dependent laboratory measurements of \citet{Matar2010} and \citet{Chaabouni2012}.

Figure~\ref{fig:porosity_sticking}(a) shows that the porosity change between validation and ISM configurations shifts $\epsilon$ by only 0.5~per~cent at the transition temperature, confirming that this configuration choice does not invalidate the cross-application of model parameters between regimes. The robustness arises because porosity primarily redistributes the spatial layout of binding sites without changing the total number per unit external surface area; the LH and ER channel rates depend on the latter quantity to first order.

Panel~(b) shows that replacing the constant $S = 0.5$ with a temperature-dependent expression motivated by the data of \citet{Matar2010} reduces the absolute warm-regime rate but preserves the qualitative temperature dependence. The shift is largest at high temperature where lower sticking reduces the effective adsorption rate. The absolute rate scaling is therefore sensitive to the sticking model, while the temperature shape (which is the principal new prediction of this work) is not. We therefore report rates with the constant-$S$ assumption in the main text and use this diagnostic as the systematic uncertainty bound on the absolute normalisation.

\bigskip
\noindent\begin{minipage}{\columnwidth}
\refstepcounter{figure}\label{fig:porosity_sticking}
\includegraphics[width=\columnwidth]{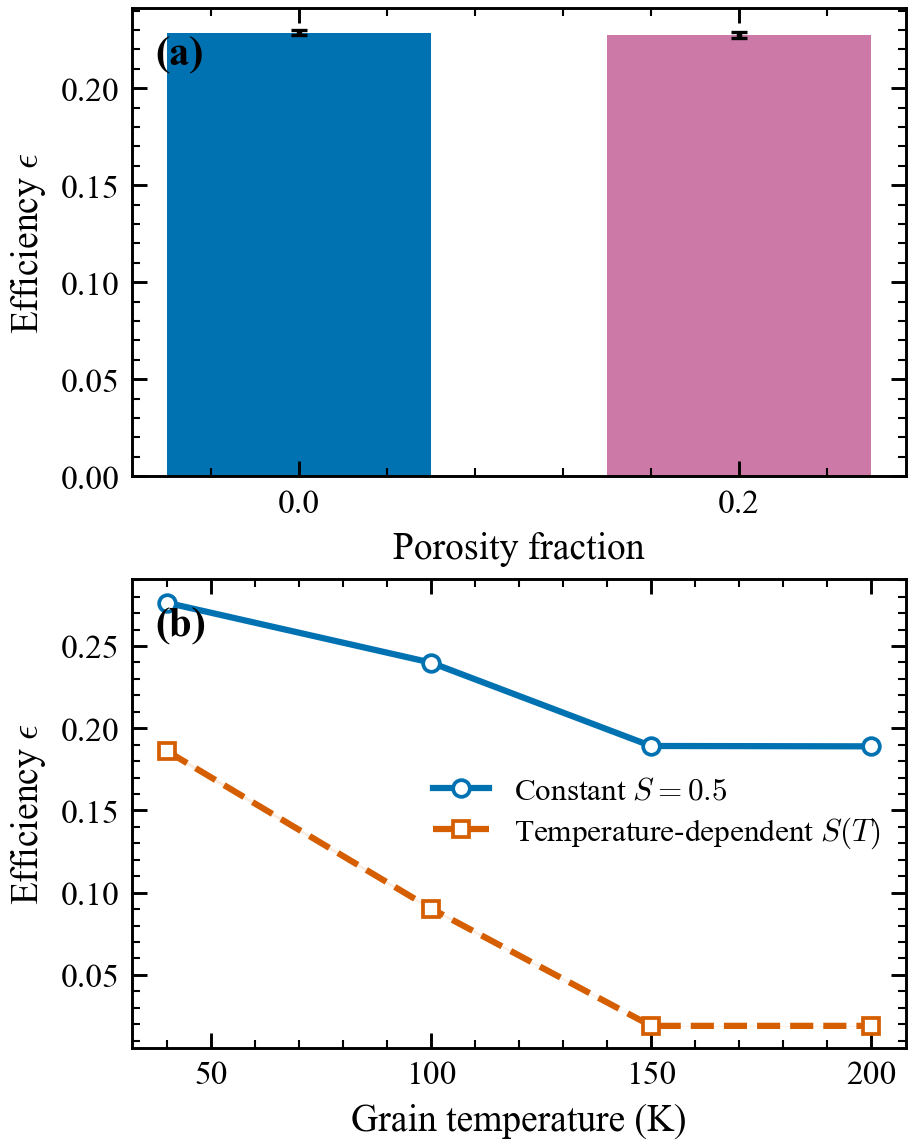}
\vskip 6pt
{\small\textbf{Figure \thefigure.} Diagnostic checks. (a) Effect of porosity on $\epsilon$ at $T = 100$~K and $\nH = 100$~\cmcube; the difference between 0.0 (validation configuration) and 0.2 (ISM configuration) is 0.5~per~cent. (b) Effect of replacing the constant sticking coefficient $S = 0.5$ with an empirical temperature-dependent law, see Eq.~\ref{eq:sticking}; the warm-regime rate is systematically lower but the qualitative temperature dependence is preserved.\par}
\end{minipage}
\medskip

\bsp
\label{lastpage}
\end{document}